\documentclass[a4paper,reqno,12pt]{amsart}
\usepackage[all]{xy}           %For commutative diagrams
\usepackage{amssymb}           %For double arrows
\usepackage{hyperref}
\usepackage{eucal}
\usepackage{epsfig}
\usepackage{pst-grad} % For gradients
\usepackage{pst-plot} % For axes
\numberwithin{equation}{section}

\newtheorem{definition}{Definition}[section]
\newtheorem{lemma}[definition]{Lemma}
\newtheorem{theorem}[definition]{Theorem}
\newtheorem{proposition}[definition]{Proposition}
\newtheorem{corollary}[definition]{Corollary}
\newtheorem{remarkth}[definition]{Remark}
\newtheorem{example}[definition]{Example}
\newenvironment{remark}{\begin{remarkth}\upshape}{\hfill$\diamond$\end{remarkth}}

\renewcommand{\emph}[1]{{\bfseries\itshape{#1}}}

\newcommand{\R}{\mathbb{R}}      %Numeros reales
\makeatletter
\newcommand\prol{\@ifstar{\@proldf}{\@prolpf}}  %% if * dual else primal
\def\@prolpf{\@ifnextchar[{\@prolpf@wrt}{\@prolpf@}}
\def\@prolpf@wrt[#1]#2{\@ifnextchar[{\@prolpf@wrt@at{#1}{#2}}{\@prolpf@wrt@{#1}{#2}}}
\def\@prolpf@wrt@at#1#2[#3]{\prolsymbol^{#1}_{#3}#2}
\def\@prolpf@wrt@#1#2{\prolsymbol^{#1}#2}
\def\@prolpf@#1{\@ifnextchar[{\@prolpf@at{#1}}{\@prolpf@@{#1}}}
\def\@prolpf@at#1[#2]{\prolsymbol_{#2}#1}
\def\@prolpf@@#1{\prolsymbol#1}
\def\@proldf{\@ifnextchar[{\@proldf@wrt}{\@proldf@}}
\def\@proldf@wrt[#1]#2{\@ifnextchar[{\@proldf@wrt@at{#1}{#2}}{\@proldf@wrt@{#1}{#2}}}
\def\@proldf@wrt@at#1#2[#3]{\prolsymbol^{*#1}_{#3}#2}
\def\@proldf@wrt@#1#2{\prolsymbol^{*#1}#2}
\def\@proldf@#1{\@ifnextchar[{\@proldf@at{#1}}{\@proldf@@{#1}}}
\def\@proldf@at#1[#2]{\prolsymbol^*_{#2}#1}
\def\@proldf@@#1{\prolsymbol^*#1}
\def\prolsymbol{\mathcal{T}}
\makeatother

\begin{document}

\title[Hamilton-Jacobi theory for singular lagrangians]{On the Hamilton-Jacobi theory for singular lagrangian systems}

\author[M. de Le\'on]{Manuel de Le\'on}
\address{Manuel de Le\'on: Instituto de Ciencias Matem\'aticas (CSIC-UAM-UC3M-UCM),
c$\backslash$ Nicol\'as Cabrera,n. 13-15, Campus Cantoblanco, UAM
28049 Madrid, Spain} \email{mdeleon@icmat.es}

\author[J. C. Marrero]{Juan Carlos Marrero}
\address{Juan C. Marrero:
Unidad asociada ULL-CSIC ``Geometr\'{\i}a Diferencial y Mec\'anica Geom\'etrica",
Departamento de Matem\'atica Fundamental,
Universidad de La Laguna, Tenerife, Canary Islands, Spain} \email{jcmarrer@ull.es}

\author[D.\ Mart\'{\i}n de Diego]{David Mart\'{\i}n de Diego}
\address{David \ Mart\'{\i}n de Diego:
Instituto de Ciencias Matem\'aticas (CSIC-UAM-UC3M-UCM),
c$\backslash$ Nicol\'as Cabrera, n. 13-15, Campus Cantoblanco, UAM
28049 Madrid, Spain}\email{david.martin@icmat.es}

\author[M. Vaquero]{Miguel Vaquero}
\address{Miguel Vaquero:
Instituto de Ciencias Matem\'aticas (CSIC-UAM-UC3M-UCM),
c$\backslash$ Nicol\'as Cabrera, n. 13-15, Campus Cantoblanco, UAM
28049 Madrid, Spain} \email{miguel.vaquero@icmat.es}

\thanks{This work has been partially supported by MICINN (Spain)
Grants  MTM2009-13383,  MTM2010-21186-C02-01 and  MTM2009-08166-E, the ICMAT Severo Ochoa project SEV-2011-0087 and the European project IRSES-project ``Geomech-246981''.
M.~Vaquero wishes to thank MICINN for a FPI-PhD Position.}
 \subjclass[2000]{}

\begin{abstract}
We develop a Hamilton-Jacobi theory for singular lagrangian systems
using the Gotay-Nester-Hinds constraint algorithm. The procedure works even if the system has secondary constraints.
\end{abstract}

\maketitle

\tableofcontents

\section{Introduction}
One of the most classical problems of theoretical mechanics is the study of constrained systems. Essentially, there are two different meanings to understand constrained systems. One refers to systems where we externally impose  constraints allowing some particular motions (external constraints). The second case is when the degeneracy of a lagrangian function imposes constraints on the phase space of the system (internal constraints). In this paper, we will restrict ourselves to this last situation.

At a first step, when the lagrangian is singular, there appear constraints restricting the admissible  positions and velocities. Later on,  the evolution of these initial  constraints may produce new constraints.

The theory of degenerate (or singular) lagrangian systems is relevant in Field theory, and just the quantization of these systems led to Dirac \cite{dirac} to develop a wonderful theory of constraints, later geometrized by Gotay, Nester and Hinds \cite{gotaythesis,gotay1,gotay0,gotay5}.
Recently, M. Leok and collaborators \cite{sosa} have studied degenerate lagrangians arising from truly mechanical systems, even in presence of additional nonholonomic constraints (see also the paper by de Le\'on and Mart{\'\i}n de Diego \cite{lmm5}).

Another important topic in theoretical mechanics is the Hamilton-Jacobi theory which allows us to  find solutions of a hamiltonian systems by means of solutions of a partial differential equation, the Hamilton-Jacobi equation. Conversely, we can treat to solve a PDE  using the characteristic curves of a Hamiltonian system (see these two standard books \cite{AM,Arnold}  for a general view of the theory and some modern approaches in terms of lagrangian submanifolds; see also \cite{rund} for a more classical view).
In \cite{lmm1,lmm3}, we have successfully extended the classical Hamilton-Jacobi theory for nonholonomic systems, and in \cite{lmm2,lmv} for classical field theories.
Therefore, it seems quite relevant to extend the Hamilton-Jacobi theory also for degenerate lagrangian systems, and this is just the goal of  the present paper.

Briefly, the standard formulation of the Hamilton-Jacobi problem is to find a function $S(t,q^A)$ (called the principal function) such that
\begin{equation}\label{hamiltonjacobi1}
\frac{\partial S}{\partial t}+H(q^A,\frac{\partial S}{\partial q^A})=0
\end{equation}

If we put $S(t,q^A)=W(q^A)-tE$, where $E$ is a constant, then $W$ satisfies
\begin{equation}\label{hamiltonjacobi2}
H(q^A,\frac{\partial W}{\partial q^A})=E;
\end{equation}
$W$ is called the characteristic function. Equations \eqref{hamiltonjacobi1} and \eqref{hamiltonjacobi2} are indistinctly referred as the Hamilton-Jacobi equations.

There have been several attempts to develop a Hamilton-Jacobi theory for degenerate lagrangian system (\cite{gomis1,gomis2,Rothe}). These procedures were based on the homogeneization of the given lagrangian, which leads to a new lagrangian system with null energy; then, it is possible to discuss the Hamilton-Jacobi equation for the constraints themselves.
The main problem is that, due to the integrability condition for the resultant partial differential equation, one can only consider first class constraints. Therefore, the treatment of the cases when second class constraints appear should be developed by {\it ad hoc} arguments (as in \cite{Rothe}, for instance). Thus, in \cite{gomis1} and \cite{gomis2} the authors only discuss the case of primary constraints.

Therefore, the Hamilton-Jacobi problem for degenerate lagrangian is far to be solved.

Our procedure to develop a geometric Hamilton-Jacobi theory is strongly inspired in two main issues. The first one in the recent approach to the Hamilton-Jacobi theory developed by Cari\~nena {\it et al} \cite{cari} (see also \cite{cari2,cari3,marmo2}, and \cite{lmm3} for the applications to nonholonomic
mechanics and field theory); and the second one, is the geometric theory of constraints due to Gotay and Nester \cite{gotay0}.

Let us recall that given an almost regular lagrangian $L:TQ\rightarrow \mathbb{R}$ one can define a presymplectic system on $M_1=FL(TQ)\subset T^*Q$, the primary constraint submanifold where $\omega_1$ is the restriction of the canonical symplectic form on $T^*Q$ to $M_1$, and $FL: {TQ}\rightarrow T^*Q$ is the Legendre transformation defined by $L$. The dynamics is obtained from the equation
\[
i_X\omega_1=dh_1,
\]
where $h_1\in C^{\infty}(M_1)$ is the projection of the energy $E_L\in C^{\infty}(TQ)$.

The above equation produces a sequence of submanifolds
\[
\cdots M_k \hookrightarrow \cdots \hookrightarrow M_2 \hookrightarrow M_ 1
\hookrightarrow T^*Q
\]
and, eventualy, a final constraint submanifold $M_f$ if the algorithm stabilize at some step.

The strategy is to consider the projection of the constraint submanifolds provided by the constraint algorithm, so that we obtain new surjective submersions onto submanifolds of the given configuration manifold. This fact permits to connect a given solution of the final constraint submanifolds $M_f$, with its projection onto $Q_f$ ($\pi_f:M_f\rightarrow Q_f$ is the surjective submersion) using a section of $\pi_f$.

The SODE problem is also discussed such that one can obtain the corresponding lagrangian picture.

We also discuss the relation of the geometric Hamilton-Jacobi problem with the Hamilton-Jacobi problem (in a traditional sense) for arbitrary extensions of $h_1$, in terms of first and second class primary and secondary constraints. Therefore, this work can be considered as the natural extension to the Hamilton-Jacobi problem of the geometrization by Gotay and Nester of the Dirac constraint algorithm.

Several examples are discussed along the paper in order to illustrate the theory.

\section{Classical Hamilton-Jacobi theory (geometric version)}

The standard formulation of the Hamilton-Jacobi problem is to find a
function $S(t, q^A)$ (called the {\bf principal function}) such that
\begin{equation}\label{hjj1}
\frac{\partial S}{\partial t} + h(q^A, \frac{\partial S}{\partial
q^A}) = 0,
\end{equation}
where $h=h(q^A, p_A)$ is the hamiltonian function of the system.
If we put $S(t, q^A) = W(q^A) - t E$, where $E$ is a constant, then
$W$ satisfies
\begin{equation}\label{hjj2}
h(q^A, \frac{\partial W}{\partial q^A}) = E;
\end{equation}
$W$ is called the {\bf characteristic function}.

Equations (\ref{hjj1}) and (\ref{hjj2}) are indistinctly referred as
the {\bf Hamilton-Jacobi equation}.

Let $Q$ be the configuration manifold, and $T^*Q$ its cotangent
bundle equipped with the canonical symplectic form
$$
\omega_Q = dq^A \wedge dp_A
$$
where $(q^A)$ are coordinates in $Q$ and $(q^A, p_A)$ are the
induced ones in $T^*Q$. In what follows,
$\pi_Q : T^*Q \longrightarrow Q$ will denote the canonical projection.

Let $h : T^*Q \longrightarrow \R$ a hamiltonian function and $X_h$
the corresponding hamiltonian vector field, say
$$
i_{X_h} \, \omega_Q = dh.
$$
Therefore, the integral curves  $(q^A(t), p_A(t))$ of $X_h$ satisfy the
Hamilton equations:
$$
\frac{dq^A}{dt} = \frac{\partial h}{\partial p_A} \, , \; \frac{dp_
A}{dt} = - \frac{\partial h}{\partial q^A}.
$$
We can define also the Poisson bracket ot two functions. Given $f$ and $g$ real functions on $T^*Q$, we define a new function $\{f,g\}$ by
\[
\{f,g\}=\omega_Q(X_f,X_g)\; ,
\]
where $X_f$ and $X_g$ are the corresponding hamiltonian vector fields.

The Poisson bracket gives us the evolution of observables, since given the hamiltonian $h$ we have
\[
\dot{f}=X_h(f)=i_{X_h}(i_{X_f}\omega_Q)=\omega_Q(X_f,X_h)=\{f,h\},
\]
and then we can rewrite the Hamilton equations as
\[
\begin{array}{l}
\dot{q}^A=\{q^A,h\}\\ \noalign{\medskip}
\dot{p}_A=\{p_A,h\}.
\end{array}
\]
Let $\lambda$ be a closed 1-form on $Q$, say $d\lambda=0$; (then,
locally $\lambda = dW$).

The following theorem gives us the relation of the Hamilton-Jacobi equation and the solutions of the Hamilton equations (see \cite{AM,Arnold}).

\begin{theorem}\label{HJTh}

The following conditions are equivalent:

\begin{enumerate}
\item[(i)] If $\sigma: I\to Q$ satisfies the equation
$$
\frac{dq^A}{dt} = \frac{\partial h}{\partial p_A},
$$
then $\lambda\circ \sigma$ is a solution of the Hamilton equations;

\item[(ii)] $d (h\circ \lambda)=0$
\end{enumerate}
\end{theorem}

We can reinterpret Theorem \ref{HJTh} as follows (see \cite{cari,lmm1,lmm3}).

Define a vector field on $Q$:
$$
X_h^{\lambda}=T\pi_Q\circ X_h\circ \lambda\; 
$$
The following diagram illustrates the construction of the vector field $X_h^\lambda$:
\[ \xymatrix{ T^*Q
\ar[dd]^{\pi_Q} \ar[rrr]^{X_h}&   & &T(T^*Q)\ar[dd]^{T\pi_Q}\\
  &  & &\\
 Q\ar@/^2pc/[uu]^{\lambda}\ar[rrr]^{X_h^{\lambda}}&  & & TQ }
\]

Then the following conditions are equivalent:
\begin{enumerate}
\item[(i)] If $\sigma: I\to Q$ satisfies the equation
$$
\frac{dq^A}{dt} = \frac{\partial h}{\partial p_A},
$$
then $\lambda\circ \sigma$ is a solution of the Hamilton equations;
\item[(i)'] If $\sigma: I\to Q$ is an integral curve of
$X_h^{\lambda}$, then $\lambda\circ \sigma$ is an integral curve of
$X_h$;
\item[(i)''] $X_h$ and $X_h^{\lambda}$ are $\lambda$-related,
i.e.
$$
T\lambda(X_h^{\lambda})=X_h \circ \lambda
$$
\end{enumerate}

Next, we have the following intrinsic version of Theorem \ref{HJTh}.

\begin{theorem}\label{HJTh2}
Let $\lambda$ be a closed $1$-form on $Q$. Then the following
conditions are equivalent:

\begin{enumerate}
\item[(i)] $X_h^\lambda$ and $X_h$ are $\lambda$-related;

\item[(ii)] $d (h\circ \lambda)=0$
\end{enumerate}
\end{theorem}

{\bf Proof:} In local coordinates, we have that
$$
X_h = \frac{\partial h}{\partial p_A} \frac{\partial}{\partial q^A} -
\frac{\partial h}{\partial q^A} \frac{\partial}{\partial p_A}
$$
and
$$
\lambda = \lambda_A(q) \, dq^A
$$
Then,
\begin{eqnarray*}
X_h^\lambda &=& \frac{\partial h}{\partial p_A}(\lambda(q)) \, \frac{\partial}{\partial q^A}\; , \\
T\lambda(X_h^\lambda) &=& \frac{\partial h}{\partial p_A} \, \frac{\partial}{\partial q^A} +
\frac{\partial h}{\partial p_A} \frac{\partial \lambda_B}{\partial q^A} \, \frac{\partial}{\partial p_B}\; ,\\
d(h \circ \lambda) &=& (\frac{\partial h}{\partial q^A} + \frac{\partial h}{\partial p_B} \frac{\partial \lambda_B}{\partial q^A} \,) dq^A\; .
\end{eqnarray*}
Since $d\lambda=0$ if and only if
$$
\frac{\partial \lambda_A}{\partial q^B} =
\frac{\partial \lambda_B}{\partial q^A},
$$
we have the equivalences between (i) and (ii).
\hfill $\Box$

\bigskip

If
$$
\lambda = \lambda_A(q) \, dq^A
$$
then the Hamilton-Jacobi equation becomes
$$
h(q^A, \lambda_A (q^B)) = const.
$$
If $\lambda = dW$ then we recover the classical formulation
$$
h(q^A, \frac{\partial W}{\partial q^A}) = const.
$$
since
$$
\lambda_A = \frac{\partial W}{\partial q^A}\; .
$$

\section{The Hamilton-Jacobi theory in the lagrangian setting}\label{lagrangian}

Let $L : TQ \longrightarrow \R$ be a lagrangian function, that is,
$$
L=L(q^A, \dot{q}^A)
$$
where $(q^A, \dot{q}^A)$ denotes the induced coordinates on the tangent bundle
$TQ$ of the configuration manifold $Q$. In what follows,
$\tau_Q : TQ \longrightarrow Q$ will denote the canonical projection.

Let us denote by
$$
S = dq^A \otimes \frac{\partial}{\partial \dot{q}^A}
$$
and
$$
\Delta = \dot{q}^A \, \frac{\partial}{\partial \dot{q}^A}
$$
the vertical endomorphism and the Liouville vector field on $TQ$ (see \cite{Leonrodrigues} for intrinsic definitions).

The Poincar\'e-Cartan 2-form is defined by
$$
\omega_L = - d\alpha_L \; , \, \alpha_L = S^*(dL)
$$
and the energy function
$$
E_L = \Delta(L)-L
$$
which in local coordinates read as
\begin{eqnarray*}
\alpha_L &=& \hat{p}_A \, dq^A \\
\omega_L &=& dq^A \wedge d\hat{p}_A\\
E_L &=& \dot{q}^A \hat{p}_A - L (q, \dot{q})
\end{eqnarray*}
where $\displaystyle{\hat{p}_A = \frac{\partial L}{\partial \dot{q}^A}}$ stand for the generalized momenta. Here $S^*$ denotes the adjoint operator of $S$.

The lagrangian $L$ is said to be regular if the Hessian matrix
$$
\left( W_{AB} = \frac{\partial^2 L}{\partial \dot{q}^A \partial
\dot{q}^B} \right)
$$
is regular, and in this case, $\omega_L$ is a symplectic form on $TQ$.

We define the Legendre transformation as a fibred mapping
$FL : TQ \longrightarrow T^*Q$ such that
$$
\langle FL(v_q), \beta_q \rangle = \langle \tilde{X}_{v_q}, \alpha_L(v_q) \rangle
$$
where $T\tau_Q(\tilde{X}_{v_q}) = v_q \in T_qQ$ and $\beta_q \in T_q^*Q$. In local coordinates we get
$$
FL(q^A, \dot{q}^A) = (q^A, \hat{p}_A),
$$
and $L$ is regular if and only if $FL$ is a local diffeomorphism.

If  $L$ is regular, then there exist a unique vector field $\xi_L$ on $TQ$
satisfying the symplectic equation
\begin{equation}
i_{\xi_L} \, \omega_L = dE_L
\end{equation}
and moreover, it automatically satisfies the second order differential equation (SODE) condition, i.e.
\[
S \, \xi_L=\Delta.
\]

If, in addition, we assume that $L$ is hyperregular, that is, $FL:TQ\rightarrow T^*Q$ is a global diffeomorphism; then we can define a (global) hamiltonian function $h:T^*Q \rightarrow \mathbb{R}$ by $h=E_L\circ FL^{-1}$. It is easy to show that $FL^*\omega_Q=\omega_L$ and that $\xi_L$ and, then $X_h$ are $FL$-related. So, the solutions of the Euler-Lagrange equations transform by $FL$ into solutions of the Hamilton equations and viceversa.
Given a vector field $Z$ on $Q$ we define a new vector field on $Q$ by
$$
\xi_L^Z = T\tau_Q \circ \xi_L \circ Z \; ,
$$
that is, we have the following commutative diagram

\[ \xymatrix{ TQ
\ar[dd]^{\tau_Q} \ar[rrr]^{\xi_L}&   & &T(TQ)\ar[dd]^{T\tau_Q}\\
  &  & &\\
 Q\ar@/^2pc/[uu]^{Z}\ar[rrr]^{\xi_L^{Z}}&  & & TQ }
\]

Now, Theorem \ref{HJTh2} can be reformulated as follows.

\begin{theorem}\label{HJTh2-bis}
Let $Z$ be a vector field on $Q$ such that $FL \circ Z$ is a closed 1-form. Then the following
conditions are equivalent:

\begin{enumerate}
\item[(i)] $\xi_L^Z$ and $\xi_L$ are $Z$-related;

\item[(ii)] $d (E_L \circ Z)=0$
\end{enumerate}
\end{theorem}

{\bf Proof:} The result follows as a direct consequence of Theorem \ref{HJTh2} and the fact that $\xi_L$ and $X_H$ are $FL$-related. \hfill $\Box$

\section{The Hamilton-Jacobi theory for singular lagrangian systems}\label{sec}

In this section we shall give a geometric approach to the Hamilton-Jacobi theory
in terms of the Gotay-Nester-Hinds constraint algorithm \cite{gotay1,gotay2}.

Let $L : TQ \longrightarrow \R$ be a singular lagrangian, that is,
the Hessian matrix
$$
\left( W_{AB} = \frac{\partial^2 L}{\partial \dot{q}^A \partial
\dot{q}^B} \right)
$$
is not regular, or, equivalently, the closed 2-form $\omega_L$ is
not symplectic.

Therefore, the equation
\begin{equation}\label{lagrangian}
i_{\xi} \, \omega_L = dE_L
\end{equation}
has no solution in general, or the solutions are not defined everywhere. Moreover the solutions do not necessarily satisfy the SODE condition. Recall that SODE condition is
\begin{equation}\label{soe}
S \, \xi=\Delta
\end{equation}
or, equivalently,
\[
T\tau_Q(X)=\tau_{TQ}(X),
\]
where $\tau_Q:TQ\rightarrow Q$ and $\tau_{TQ}:TTQ\rightarrow TQ$ are the canonical projections.

Singular lagrangian system have been extensively studied by P.A.M. Dirac and P. Bergmann (see{dirac}), in order to obtain a  procedure for  canonical quantization of local gauge theories. They developed an algorithm (called Dirac-Bergmann theory of constraints) that has been later geometrized  by M.J. Gotay and J. Nester  \cite{dirac,gotay1,gotay2}.

In the sequel,  we will assume that $L$ is almost regular, which means that:

\begin{itemize}

\item
$M_1= FL(TQ)$ is a submanifold of $T^*Q$;

\item The restriction of the Legendre mapping $FL_1 : TQ \longrightarrow M_1$ is a
submersion with connected fibers.
\end{itemize}

In this case, $M_1$ is called the submanifold of primary constraints.

If $L$ is almost regular, since $\ker( TFL)=\ker( \omega_L)\cap V(TQ)$, where
$V(TQ)$ denotes the vertical bundle, and the fibers are connected then a direct computation shows that
$E_L$ projects onto a function
$$
h_1 : M_1 \longrightarrow \R\; .
$$

Denote by $j_1 : M_1 \longrightarrow T^*Q$ the natural inclusion and
define
$$
\omega_1 = j_1^* (\omega_Q) \; .
$$

Consider now the equation
\begin{equation}\label{singular}
i_X \, \omega_1 = dh_1\; .
\end{equation}

There are two possibilities:
\begin{itemize}
\item There is a solution $X$ defined at all the points of $M_1$;
such $X$ is called a global dynamics and it is a solution (modulo
$\ker \, \omega_1$). In other words, there are only primary
constraints.

\item Otherwise, we select the submanifold $M_2$ formed by those
points of $M_1$ where a solution exists. But such a solution $X$ is
not necessarily tangent to $M_2$, so we have to impose an additional tangency
condition, and we obtain a new submanifold $M_3$ along which there
exists a solution. Continuing this process, we obtain a sequence of
submanifolds
\[
\cdots M_k \hookrightarrow \cdots \hookrightarrow M_2 \hookrightarrow M_ 1
\hookrightarrow T^*Q
\]
where the general description of $M_{l+1}$ is
\[
M_{l+1}:=\{p\in M_{l} \textrm{ such that there exists } X_p\in T_pM_l \\ 
\textrm{ satisfying } i_X\omega_1=dh_1 \}.
\]
If the algorithm stabilizes at some $k$, say $M_{k+1}=M_k$, then we
say that $M_k$ is the final constraint submanifold which is denoted by $M_f$, and then there
exists a well-defined solution $X$ of (\ref{singular}) along $M_f$.
\end{itemize}
\begin{remark}
There is another characterization of the submanifolds $M_l$ that we will describe now. If $N$ is a submanifold of $M_1$ then we define
\[
TN^{\perp}=\{Z\in T_p(M_1), \ p\in N \textrm{ such that } \omega_1(X,Z)=0 \textrm{ for all } X\in T_pN\}.
\]

Then, at any point $p\in M_l$ there exists $X_p\in T_pM_l$ verifying $i_X\omega_1=dh_1$ if and only if $\langle TM_k^{\perp},dh_1 \rangle=0$, (see \cite{gotay0}).

Hence, we can define the $l+1$ step of the constraint algorithm as
\[
M_{l+1}:=\{p\in M_l\textrm{ such that }\langle TM_l^{\perp},dh_1 \rangle(p)=0\} \; ,
\]
where ${TM_l}^{\perp}$ is defined as above.
\end{remark}
\subsection{Case I: There is a global dynamics}

In this case there exists a vector field $X$ on $M_1$ such that
$$
(i_X \, \omega_1 = dh_1)|_{M_1}.
$$

Moreover, we have $\pi_1(M_1)=Q$, where $\pi_1$ is the restriction
to $M_1$ of the canonical projection $\pi_Q : T^*Q \longrightarrow Q$.

Next, assume that $\gamma$ is a closed 1-form on $Q$ such that $\gamma(Q)
\subset M_1$. Define now a vector field $X^\gamma$ on $Q$ by putting
$$
X^\gamma = T\pi_1 \circ  X \circ \gamma\; .
$$
The following diagram summarizes the above construction:
\[
\xymatrix{
M_1\ar[dd]^{\pi _1} \ar[rrr]^{X}&   & &TM_1\ar[dd]^{T\pi _1}\\
  &  & &\\
 Q\ar@/^2pc/[uu]^{\gamma_1}\ar[rrr]^{X^\gamma}&  & & TQ }
\]

Here $\gamma_1$ denotes the restriction of $\gamma$.

We have
\begin{eqnarray*}
\gamma_1^*(i_{X-T\gamma_1(X^\gamma)} \, \omega_1) &=& \gamma_1^*(i_X \,
\omega_1) - \gamma_1^*(i_{T\gamma_1(X^\gamma)} \, \omega_1) \\
&=& \gamma_1^* d(h_1) - \gamma_1^*(i_{T\gamma_1(X^\gamma)} \, \omega_1)\\
&=&d(h_1 \circ \gamma_1)
\end{eqnarray*}
since $\gamma^*(i_{T\gamma_1 X^\gamma} \, \omega_1) = i_{X^\gamma} \,
(\gamma_1^* \omega_1)  = 0$, because
$$
\gamma_1^* \omega_1 = \gamma_1^* j_1^* \omega_Q = (j_1 \circ \gamma_1)^* \omega_Q = \gamma^* \omega_Q =  - d\gamma = 0
$$

Therefore, taking into account that  $V\pi_1\oplus T\gamma_1(TQ)= TM_1$ and $\omega_1$ (as it happens with $\omega_Q$)
vanishes acting on two vertical tangent vectors with respect to the canonical projection $\pi_1:M_1
\rightarrow Q$, we deduce the following:
\begin{equation}\label{condition-1}
X-T\gamma_1(X^\gamma) \in \ker (\omega_1) \Leftrightarrow d(h_1 \circ
\gamma_1) = 0
\end{equation}

Moreover, we will show that it is possible to refine condition (\ref{condition-1}) and to prove that   $X$ and $X^{\gamma}$ are $\gamma_1$ related.

First at all, it is clear that for any point $p$ of $M_1$
\[
T_p(T^*Q)=T_pM_1+V_p(T^*Q)
\]
where $V(T^*Q)$ denotes the space of vertical tangent  vectors at $p$.

In addition, $X-T\gamma_1(X^{\gamma})$ is vertical at the points of $\textrm{Im}(\gamma_1)$, so given any $Z\in V_p(T^*Q)$, $p\in \textrm{Im}(\gamma_1)$, we deduce
\[
\omega_Q(X-T\gamma_1(X^{\gamma}),Z)=0 \hbox{ along $\textrm{Im} ( \gamma_1)$}
\]
since $\omega_Q$ vanishes on two vertical tangent vectors.

Now, given $Z\in T_pM_1$ we have
\[
\omega_Q(X-T\gamma_1(X^{\gamma}),Z)=\omega_1(X-T\gamma_1(X^{\gamma}),Z)=0
\]
because $X-T\gamma_1(X^{\gamma})\in \ker( \omega_1)$, and we obtain that  $\omega_Q(X-T\gamma_1(X^{\gamma}),Z)$ $=0$ for any tangent vector $Z\in T_p(T^*Q)$ on any point of $\textrm{Im} ( \gamma_1)$. Since $\omega_Q$ is non-degenerate we deduce that $X=T\gamma_1(X^{\gamma})$ on $\textrm{Im} ( \gamma_1)$.

In conclusion, we have the following result
\begin{proposition}\label{hj111}
\[
X \textrm{ and }T\gamma_1(X^\gamma)  \textrm{ are $\gamma_1$ related } \Leftrightarrow d(h_1 \circ
\gamma_1) = 0
\]
\end{proposition}

\begin{remark}{\rm
As a consequence of the above result, if $h_1$ is constant along $\gamma_1(Q)$ then $\gamma_1$ maps the integral curves of $X^\gamma$
on integral curves of $X$. So $d(h_1\circ \gamma_1)=0$ can be considered as the Hamilton-Jacobi equation in this case.
}
\end{remark}

\subsection{Case II: There are secondary constraints}

In this case, the algorithm produces a sequence of submanifolds as follows
\[
\cdots M_k \hookrightarrow \cdots\hookrightarrow M_2 \hookrightarrow M_ 1
\hookrightarrow T^*Q
\]

We assume that the projections $Q_r:=\pi_Q(M_r)$ are submanifolds, and that the corresponding projections $\pi_r:M_r\rightarrow Q_r$ are fibrations, where $\pi_r$ is the restriction of $\pi_Q$ to $M_r$.

The constraint algorithm produces a solution $X$ of the equation
$$
(i_X \, \omega_1 = dh_1)_{|M_f} \; ,
$$
where $X$ is a vector field on $M_f$.

Coming back to the Gotay-Nester-Dirac algorithm we can summarize the situation in the following diagram:
\[
\xymatrix{
TQ \ar@{->}[r]^{FL} \ar[dr]_{FL_1} & T^*Q \ar[dr]^{\pi _Q} &   \\
  &  M_1 \ar@{_{(}->}[u] \ar[r]^{\pi _1} & Q \\
  &  M_2 \ar@{_{(}->}[u] \ar[r]^{\pi _2 }&\ar@{_{(}->}[u] Q_2  \\
  &      \ar@{_{(}->}[u] & \ar@{_{(}->}[u]  \\
  &      \ar@{.}[u] &\ar@{.}[u]   \\
  &  M_f \ar@{_{(}->}[u] \ar[r]^{\pi _f } & \ar@{_{(}->}[u] Q_f
      }
\]

Assume now that $\gamma$ is a closed 1-form on $Q$ such that
\begin{itemize}
\item $\gamma(Q) \subset M_1$.
\item $\gamma(Q_f) \subset M_f$.
\end{itemize}
As in Case I,  $\gamma$ allows us to define a vector field $X^\gamma\in \mathfrak X (Q_f)$ by
$$
X^{\gamma} = T\pi_f \circ  X \circ \gamma_f\; .
$$
\[
\xymatrix{
M_f\ar[dd]^{\pi _1} \ar[rrr]^{X}&   & &TM_f\ar[dd]^{T\pi _f}\\
  &  & &\\
Q_f\ar@/^2pc/[uu]^{\gamma_f}\ar[rrr]^{X^\gamma}&  & & TQ_f
}
\]
Here $\gamma_f$ is the restriction of $\gamma$ to $Q_f$.

Now, given $q\in Q_f$, we have
\[
\begin{array}{l}
i_{\left(X(\gamma_1(q))-T_{q}\gamma_f(X^{\gamma}(q))\right)}\, \omega_1\circ T_{q}\gamma_1= i_{X(\gamma(q))}\, \omega_1\circ T_{q}\gamma_1-i_{T_{q}\gamma_f(X^{\gamma}(q))}\, \omega_1\circ T_{q}\gamma_1\\ \noalign{\medskip}
=dh_1({\gamma_f(q)})\circ T_{q}\gamma=d(h_1\circ \gamma_1)(q)\; .
\end{array}
\]
Observe that  since $\gamma_f$ is the restriction of $\gamma_1$ we have $T_{q}\gamma_f(X^{\gamma}(q))=T_{q}\gamma(X^{\gamma}(q))$. Therefore, given $Y_{q}\in T_{q}Q$ then $T_{q}\gamma_1(Y(q))=T_{q}\gamma(Y(q))$, and we deduce that
\[
\begin{array}{l}
i_{T_{q}\gamma_f(X^{\gamma}(q))}\, \omega_1\circ T_{q}\gamma(Y(q))= \omega_1(T_{q}\gamma(X^{\gamma}(q)), T_{q}\gamma(Y(q)))\\ \noalign{\medskip}
=(\gamma_1^*\omega_1)(X^{\gamma}(q),Y(q))=\gamma_1^*j_1^*\omega_Q(X^{\gamma}(q),Y(q))=d\gamma(X^{\gamma}(q),Y(q))=0\; .
\end{array}
\]
The previous discussion can be applied to every point $q\in Q_f$; therefore, taking into account that $\omega_1$  vanishes acting on two vertical tangent vectors and $V\pi_1\oplus T\gamma_1(TQ)= TM_1$, we can deduce the following
\[
X-T\gamma_f(X^\gamma) \mathop{\in}_{M_f} \ker( \omega_1) \Leftrightarrow d(h_1 \circ
\gamma_1)_{|Q_f} = 0
\]

Usin g a similar argument that in Case I, it is possible to deduce that  $X$ and $X^{\gamma}$ are $\gamma_f$ related since we have
\[
T_p(T^*Q)=T_pM_1+V_p(T^*Q)
\]
for all $p\in M_f$.

Therefore, we deduce the following.

\begin{proposition}\label{hj111}
\[
X \textrm{ and }T\gamma_1(X^\gamma)  \textrm{ are $\gamma_f$ related } \Leftrightarrow d(h_1 \circ
\gamma_1)_{|Q_f} = 0
\]
\end{proposition}

\begin{remark}{\rm
Notice that the condition $X-T\gamma_f(X^\gamma)\in\textrm{Ker}(\omega_f)$ along $\gamma_f(Q_f)$
implies that, if $\sigma : \R \longrightarrow Q_f$ is an integral curve of $X^\gamma$,
then $\sigma_\gamma = \gamma \circ \sigma : \R \longrightarrow M_f$ is an integral curve of $X$

Therefore, the condition
$$
d(h_1 \circ \gamma_1)_{|Q_f} = 0
$$
could be still considered as the Hamilton-Jacobi equation in this
context.
}
\end{remark}

\subsection{Hamilton-Jacobi theory for further geometric constraint equations}

Besides of the equation $i_X \, \omega=dh_1$ introduced in the previous section, other equations have been studied in the literature starting with the same data, that is, a singular lagrangian. For completeness, in this section we will discuss some of these equation of motions.

A good reference for these  topics is \cite{gotaythesis}.

\subsubsection{Extended equation of Motion and the Dirac conjeture}\label{DIRAC-a}
A constraint is called first class provided its Poisson bracket with every other constraint
weakly vanishes, and second class otherwise (see Section \ref{dirac} for more details). Dirac \cite{dirac} conjectured that
all first-class secondary constraints generate `gauge transformations' which leave the
physical state invariant. See, for instance \cite{gotay4} and references therein for the discussion about the avaibility of the Dirac conjeture. Moreover, the motivation of our study will be more clear in section \ref{dirac}.

Withour entering in physical discussions, we will analyze if it is possible to extend our Hamilton-Jacobi formalism for the equations derived assuming Dirac conjeture. Therefore, we need first to discuss the geometry of this `extended equation' for singular lagrangians.

Suppose that we are in the conditions of the previous section. We have $j_1:M_1\rightarrow T^*Q$ where $M_1$ is a submanifold and $j_1$ the inclusion, and a constrained hamiltonian $h_1:M_1\rightarrow \mathbb{R}$. As before, we study the presymplectic system $(M_1, \omega_1, dh_1)$ and apply the Gotay-Nester-Hinds algorithm, assuming that we reach to a  final constraint manifold $M_f$. Denote by $j_f:M_f\rightarrow M_1$ the inclusion. Now we say that a vector field $X$ on $M_f$ is a  solution of the extended equations of motion if $X$ can be writen
\begin{equation}\label{extended}
X=Y+Z
\end{equation}
where $Y$ and $Z$ are vector fields on $M_f$, such that $i_Y \, \omega_1=dh_1$ and $Z\in\ker(\omega_f)$ where $\omega_f:=j_f^*(\omega_1)$.

We can now obtain a less restrictive version of the previous Hamilton-Jacobi theory, which gives solutions of the extended equations of motion.

Assume again that $\gamma$ is a $1$-form on $Q$ such that
\begin{enumerate}
\item Im$(\gamma)\subset M_1$
\item Im$(\gamma_f)\subset M_f$
\item $d\gamma=0$
\end{enumerate}

From a fixed solution $X$ of the extended equation, we can define $X^{\gamma} = T\pi_f \circ  X \circ \gamma_f$.

Proceeding as in the previous section, we have
\begin{eqnarray*}
\gamma_f^*(i_{X-T\gamma_f(X^\gamma)} \, \omega_f) &=& \gamma_f^*(i_X
\, \omega_f) - \gamma_f^*(i_{T\gamma_f(X^\gamma)} \, \omega_f) \\
&=& \gamma_f^* dh_1 - \gamma_f^*(i_{T\gamma (X^\gamma)} \, \omega_f)\\
&=&d(h_1 \circ \gamma_f)\; .
\end{eqnarray*}
since $\gamma$ is closed.

Using  similar arguments that in the previous section, we deduce the following
\begin{proposition}\label{112}
Under the above conditions, we have\hfill
$$X-T\gamma_f(X^\gamma)\mathop{\in}_{M_f} \ker(\omega_f)\Leftrightarrow d(h_1\circ\gamma_f)=0.$$
\item If $d(h_1\circ\gamma_f)=0$, then $X-T\gamma_f(X^\gamma)\in\ker (\omega_f)$.
 \end{proposition}

{\bf Proof:}
It follows the same lines of the proofs of Proposition \ref{hj111} but now observing that
\[
TM_f=T\gamma_f(TQ_f)\oplus V\pi_f
\]
and $V\pi_f\subset V_{M_f}\pi_Q$.
 \hfill $\Box$.

By the last proposition $T\gamma_f(X^\gamma)=X+\tilde{Z}$, with $\tilde{Z}\in \ker \omega_f$. Then,
form (\ref{extended}) we have that $T\gamma_f(X^\gamma)=Y+(Z+\tilde{Z})$.
So,  $T\gamma_f(X^\gamma)$  is a solution of the extended equations of motion.

Therefore, the condition
$$
d(h_1 \circ \gamma_f) = 0
$$
could be still considered as the Hamilton-Jacobi equation in this
context.

\subsubsection{Hinds algorithm}\label{HINDS-a}

Besides of the Gotay-Nester-Hinds algorithm, other approaches have been discussed in the literature. In particular we briefly recall the algorithm introduced by Hinds (see Gotay \cite{gotay0} for a detailed discussion).
Hinds algorithm also start considering the equation $i_X\omega_1=dh_1$ as the Gotay-Nester-Hinds algorithm. The algorithm
generates a descending sequence of constraint submanifolds. In the favorable case, the
algorithm stabilizes at a final constraint submanifold which we will denote again
by $N_f$ (see discussion below). It is important to point out that, in general, this constraint submanifold $N_f$ will be different from the final
constraint submanifold obtained by the Gotay-Nester-Hinds algorithm, that is $N_f\neq M_f$.  In principle, both
algorithms start to diverge from each other after the second step.

In more geometric terms, assume that we are in the conditions of the previous section. Define $N_1:=M_1$ as we did before and denotes $N_{l+1}$ for $l>1$ the following subset
\[
N_{l+1}:=\{p\in N_l \textrm{ such that exists } X\in T_pN_l \textrm{ verifying }i_{X}\omega_l=dh_l\},
\]
where, if we call $k_l:N_l\rightarrow N_1$ the natural inclusion, then $\omega_l:=k_l^*\omega_1$ and $h_l:=k_l^*h_1$. We obtain the sequence of submanifolds
$$
\cdots N_k \hookrightarrow \cdots \hookrightarrow N_2 \hookrightarrow N_ 1=M_1
\hookrightarrow T^*Q.
$$
Again if the algorithm stabilizes, i.e. $N_k=N_{k+1}$, then we say that $N_k$ is the final constraint manifold, $N_f$. In this case, the Hinds algorithm produces a solution $X\in {\mathfrak X}(N_f)$ of the equation
\[
i_{X}\omega_f=dh_f.
\]
This equation is less restrictive than \eqref{singular}, and so the two algorithms diverge for $l\geq 2 $. We will come back later to the above equation.

Now, we can  develop a Hamilton-Jacobi theory in this setting.

Assume that there exists a $1$-form $\gamma$ on $Q$ satisfying
\begin{enumerate}
\item Im$(\gamma_f)\subset N_f$\, ,
\item $d\gamma=0$ along $N_f$.
\end{enumerate}
Then we can define $X^{\gamma} = T\pi_f \circ  X \circ \gamma_f$ and state the equivalent Hamilton-Jacobi theory. 
The proof follws the same lines that in proposition \ref{112}.

\begin{proposition}\label{113}
\[
 X-T\gamma_f(X^\gamma)\in \ker(\omega_f)\Leftrightarrow d(h_1\circ\gamma_f)=0.
 \]
\end{proposition}

\subsection{Relation to the Dirac-Bergmann theory of constraints}\label{dirac}
In this section we will discuss the relation of the Gotay and Nester theory with the original Dirac-Bergmann theory of constraints.

Assume that we begin with an almost regular lagrangian $L:TQ\rightarrow \mathbb{R}$. Then there exists an open neighbourhood, $U\subset T^*Q$ where in canonical coordinates $(q^A,p_A)$, $M_1\cap U$ is given by the vanishing of functions $\Phi^i(q^A,p_A)$ defined on $U$. The functions $\Phi^i$ are called primary constraints.

Remember that we can project $E_L$ to $h_1:M_1\rightarrow \mathbb{R}$, and any extension of $h_1$ to $U$ should be of the form
\[
H=h+u_i\Phi^i \; ,
\]
where $h$ is an arbitrary extension to $U$ of $h_1$. The functions $u_i$, $1\leq i\leq 2\dim Q-\dim M_1$ are Lagrange multipliers to be determined.

Acording to Dirac the equations of motion are
\[
\begin{array}{l}
\dot{q}^A=\frac{\partial H}{\partial p_A}+u_i\frac{\partial \Phi^i}{\partial  p_A} \\ \noalign{\medskip}
\dot{p}_A=-\frac{\partial  H}{\partial  q^A}-u_i\frac{\partial  \Phi^i}{\partial  q^A}
\end{array}
\]
which must hold over $U_1:=M_1\cap U$. If we denote $j_1:U_1\rightarrow U$ the inclusion, and $\omega_1=j_1^*\omega_Q$,  the preceding equations  can be equivalen rewritten as
\begin{equation}\label{ecu}\begin{array}{lcr}
 i_X\omega_1=dh_1 &\textrm{or}& (i_X\omega_Q=dh+u_id\Phi^i)_{|U_1}
\end{array}
\end{equation}
which are the equations that we have considered in the Gotay-Nester-Hinds algorithm.

Since $X$ must be tangent to $U_1$ we should have
\[
\begin{array}{lll}
0&=&(X(\Phi^i))_{|U_1}=\{\Phi^i,H\}_{|U_1}=\{\Phi^i,h+u_j\Phi^j\}_{|U_1} \\ \noalign{\medskip}
&=&(\{\Phi^i,h\}+u_j\{\Phi^i,\Phi^j\})_{|U_1}.
\end{array}
\]

These equations can be trivially satified,  determine some Lagrange multipliers or add new constraints on the variables $q^A$, $p_A$ over $U_1$. These new constraints, if any, are called secondary constraints. Suppose that we have obtained the secondary constraints $\xi^{\alpha}$. So, we have to restrict the dynamics to $U_2:=U_1\cap (\xi^{\alpha})^{-1}\{0\}$.

Again, the solution must be tangent to $U_2$ and it requires that
\[
\begin{array}{lll}
0&=&(X(\xi^{\alpha}))_{|U_2}=\{\xi^{\alpha},H\}_{|U_2}=\{\xi^{\alpha},h+u_i\Phi^i\}_{|U_2} \\ \noalign{\medskip}
&=&(\{\xi^{\alpha},h\}+u_i\{\xi^{\alpha},\Phi^i\})_{|U_2}.
\end{array}
\]
As before, these equations may determine more Lagrange multipliers or add new  constraints  to the picture, that is, 
new secondary constraints. Iterating this procedure, if the algorithm stabilizes, we arrive to a set $U_f$ which is an open subset of the final constraint manifold $M_f$ obtained by the Gotay-Nester-Hinds algorithm (see \cite{gotay0} for a proof ).

It is necesary to introduce some definitions. We say that a function defined on $U$ is \textbf{first class} if its Poisson bracket with every constraint (primary and secondary) vanishes. Otherwise, it is said to be of \textbf{second class}.

We can reorder constraints into first class or second class.  We will denote by $\chi^a$ and $\xi^b$, the primary first and second class constraints, respectively; and by $\psi^c$ and $\theta^d$,  the secondary first and second class constraints, respectively. We will also denote by $\mu_a,\lambda_b$ the corresponding Lagrange multipliers for the primary first and second class constraints, respectively.

So, if the problem has a solution, we must obtain a vector field $X$ over $U_f$, which satisfies the equations
\[
(i_X \, \omega_Q = dh+\mu_ad\chi^a+\lambda_bd\xi^b)_{|M_f}.
\]
The $\lambda_b$'s are determined functions and the $\mu_a$'s can be varied to obtain other admissible solutions. In consequence, it is also clear that primary first class constraints correspond to gauge transformations which leave the physical state invariant. As we have discussed before, Dirac conjectured that the first class secondary constraints may also generate gauge transformations, therefore, the generalized  equations of motion discussed in Subsection \ref{DIRAC-a} are locally rewritten as
\begin{equation}\label{dirac1}
(i_X \, \omega_Q = dh+\mu_ad\chi^a+\lambda_bd\xi^b+v_cd\psi^c)_{|M_f}.
\end{equation}
where $\lambda_b$ are still determined functions and $\mu_a$ and $v_c$ can be varied arbitrarily. The hamiltonian $h+\mu_a\chi^a+\lambda_b\xi^b+v_c\psi^c$ is called the extended hamiltonian, and equation \eqref{dirac1} the extended equation of motion following the notation of \cite{gotaythesis}. Geometrically the solutions of \eqref{dirac1} are just
\[
X=Y+Z,
\]
where $Y$ is a vector field on $M_f$ solution of the equations of motion, \ref{ecu}, and $Z\in \ker(\omega_f)$ where $\omega_f$ is the restriction  of $\omega_1$ to $M_f$.

\begin{remark}
If we proceed in the same way with the Hinds algorithm developed in Subsection \ref{HINDS-a}, we will arrive to solutions $X$ satisfying
\[
(i_X \, \omega_Q = dh+\mu_ad\chi^a+\lambda_bd\xi^b +\overline{v}_{\overline{c}}\overline{\psi}^{\overline{c}}+\overline{w}_{\overline{d}}\overline{\theta}^{\overline{d}})_{|N_f},
\]
where $\overline{v}_{\overline{c}}$, $\overline{w}_{\overline{d}}$ are the Lagrange multipliers corresponding to the constraints $\overline{\psi}^{\overline{c}}$ and $\overline{\theta}^{\overline{d}}$. Note that $\overline{\psi}^{\overline{c}}$ and $\overline{\theta}^{\overline{d}}$ now correspond to the secondary constraints of the final constraint manifold $N_f$ in the Hinds algorithm.
\end{remark}
\subsection{Examples}

Now we illustrate the previous propositions with several examples.

\subsubsection{There are only primary constraints}

\begin{example}{\rm
This example is discussed by O. Krupkova in \cite{Krup1}. Let $L$ be the Lagrangian $L:T{\mathbb{R}}^3\rightarrow \mathbb{R} $  given by
\[
L(q^1,q^2,q^3,\dot{q}^1,\dot{q}^2,\dot{q}^3)=\frac{1}{2}(\dot{q}^1+\dot{q}^2)^2.
\]
Then $FL$ is given by $FL:T{\mathbb{R}}^3\rightarrow T^*{\mathbb{R}}^3$
\[
FL(q^1,q^2,q^3,\dot{q}^1,\dot{q}^2,\dot{q}^3)=(q^1,q^2,q^3,\dot{q}^1+\dot{q}^2,\dot{q}^1+\dot{q}^2,0),
\]
and the primary constraints are
\[
\begin{array}{lr}
\Phi^1(q^A,p_A)=p_1-\,p_2 &  \Phi^2(q^A,p_A)=p_3.
\end{array}
\]
So
\[
M_1=\{(q^1,q^2,q^3,p_1,p_2,p_3)\in\mathbb{R}^6 \textrm{ such that } p_1=p_2, \ p_3=0\}
\]
and we can use $(q^1,q^2,q^3,p_1)$ as coordinates on $M_1$.

It follows that
\[\begin{array}{l}
E_L=(\dot{q}^1+\dot{q}^2)\dot{q}^1+(\dot{q}^1+\dot{q}^2)\dot{q}^2-L=L \\ \noalign{\medskip}
h_1=\frac{1}{2}({p_1})^2  \\ \noalign{\medskip}
\omega_1=dq^1\wedge dp_1+dq^2\wedge dp_1  \\ \noalign{\medskip}
\textrm{Ker}(\omega_1)=\displaystyle{\left\{\frac{\partial}{\partial q^3},\frac{\partial}{\partial q^1}-\frac{\partial}{\partial q^2} \right\}} \\
\end{array}
\]
and a particular extension of the hamiltonian is
\[
h(q^A,p_A)=\frac{1}{2}({p_1})^2
\]

It is easy to see that at the points of $M_1:=\textrm{Im}(FL)$
\[
\{\Phi^i, h +u^1\Phi^1+u^2\Phi^2\}=0, \ i=1,2.
\]

So we have global dynamics on $M_1$ and it holds that
\[
\{\Phi^1,\Phi^2\}=0
\]
and we conclude that there are only first class constraints.
The solutions of $ (i_X\omega_1=dh_1)_{|M_1}$ on $M_1$ are given by \[X=p_1\frac{\partial}{\partial q^1}+f_1\frac{\partial}{\partial q^3}+f_2(\frac{\partial}{\partial q^1}-\frac{\partial}{\partial q^2}),\] where $f_1$ and $f_2$ are functions on $M_1$.

We now look for $\gamma\in \Lambda^1(Q)$ such that
\begin{enumerate}
\item $\gamma(Q)\subset M_1$
\item $d(h_1\circ \gamma)=0$
\item $d\gamma=0$
\end{enumerate}

Suppose \[\gamma(q^1,q^2,q^3)=(q^1,q^2,q^3,\gamma_1(q^1,q^2,q^3),\gamma_2(q^1,q^2,q^3),\gamma_3(q^1,q^2,q^3)),\] then, $\gamma(Q)\subset M_1$ implies that $\gamma_1=\gamma_2$ and $\gamma_3=0$.

The condition $d(h_1\circ \gamma)=0$ implies that $\frac{1}{2}({\gamma_1})^2=\textrm{constant}$, and because of that $\gamma_1=c$ where $c$ is a constant and so $\gamma_2=c$.

 Now $\gamma(q^1,q^2,q^3)=(q^1,q^2,q^3,c,c,0)$ and  $d\gamma=0$ is trivially satisfied.

If we take the general solution $p_1\frac{\partial}{\partial q^1}+f_1\frac{\partial}{\partial q^3}+f_2(\frac{\partial}{\partial q^1}-\frac{\partial}{\partial q^2})$ then we obtain $X^{\gamma}=T\pi_q\circ X\circ \gamma=c\frac{\partial}{\partial q^1}+(f_1\circ \gamma)\frac{\partial}{\partial q^3}+(f_2\circ\gamma)(\frac{\partial}{\partial q^1}-\frac{\partial}{\partial q^2})$. If we apply $T\gamma(q^A)(X^{\gamma}(q^A))=X(\gamma(q^A))$, then we recover the solution $X$ over the points of $\gamma$.
It is clear, that integral curves of $X^{\gamma}$ are applied by $\gamma$ into integral curves of $X$ along ${\hbox{Im }\gamma}$.
}
\end{example}

\begin{example}{\rm
This example has been discussed by J. Barcelos-Neto and N.R.F. Braga \cite{barcelos}. Let $L$ be the Lagrangian $L:T{\mathbb{R}}^4\rightarrow \mathbb{R} $  given by
\[
L(q^1,q^2,q^3,q^4,\dot{q}^1,\dot{q}^2,\dot{q}^3,\dot{q}^4)=(q^2+q^3)\dot{q}^1 +q^4\dot{q}^3+\frac{1}{2}\left((q^4)^2-2q^2q^3-(q^3)^2\right).
\]

Then $FL$ is given by $FL:T{\mathbb{R}}^4\rightarrow T^*{\mathbb{R}}^4$
\[
FL(q^1,q^2,q^3,q^4,\dot{q}^1,\dot{q}^2,\dot{q}^3,\dot{q}^4)=(q^1,q^2,q^3,q^4,q^2+q^3,0,q^4,0)
\]
and the primary constraints are
\[\begin{array}{lr}
\Phi^1(q^A,p_A)=p_1-q^2-q^3 , & \Phi^2(q^A,p_A)=p_2 , \\ \noalign{\medskip} \Phi^3(q^A,p_A)=p_3-q^4 , & \Phi^4(q^A,p_A)=p_4.
\end{array}
\]
So
\[
\begin{array}{l}
M_1=\{(q^1,q^2,q^3,q^4,p_1,p_2,p_3,p_4)\in\mathbb{R}^8 \textrm{ such that }\\ \noalign{\medskip}%\\ \noalign \medskip
\quad \quad \quad \quad p_1=q^2+q^3, \ p_2=0, \ p_3=q^4, \ p_4=0\}.
\end{array}
\]
and we can use $(q^1,q^2,q^3,q^4)$ as coordinates on $M_1$.

It follows that
\[\begin{array}{l}
E_L=(q^2+q^3)\dot{q}^1+q^4\dot{q}^3-L=-\frac{1}{2}((q^4)^2-2q^2q^3-(q^3)^2) \\  \noalign{\medskip}
h_1=-\frac{1}{2}((q^4)^2-2q^2q^3-(q^3)^2) \\ \noalign{\medskip}
\omega_1=dq^1\wedge dq^2+dq^1\wedge dq^3+ dq^3\wedge dq^4  \\ \noalign{\medskip}
\textrm{Ker}(\omega_1)=\{0\} \\
\end{array}
\]
so $(M_1, \omega_1)$ is a symplectic manifold.

It we prefer to follow the Dirac-Bergmann algorithm, then one should take an extension
$
h(q^A,p_A)=-\frac{1}{2}((q^4)^2-2q^2q^3-(q^3)^2)
$ of $h_1$.
 It is easy to see that at the points of $M_1:=\textrm{Im}(FL)$
\[\begin{array}{l}
\{\Phi^1, h +u^1\Phi^1+u^2\Phi^2+u^3\Phi^3+u^4\Phi^4\}=-u^2-u^3\\\noalign{\medskip}
\{\Phi^2, h +u^1\Phi^1+u^2\Phi^2+u^3\Phi^3+u^4\Phi^4\}=-q^3+u^1\\\noalign{\medskip}
\{\Phi^3, h +u^1\Phi^1+u^2\Phi^2+u^3\Phi^3+u^4\Phi^4\}=-q^2-q^3+u^1-u^4 \\\noalign{\medskip}
\{\Phi^4, h +u^1\Phi^1+u^2\Phi^2+u^3\Phi^3+u^4\Phi^4\}= u^3+u^4,
\end{array}
\]
which determine completely the Lagrange multipliers:
\[
u^1=q^3\, , u^2=q^4\, ,
u^3=-q^4\, ,
u^4=-q^2,
\]
and then all the constraints are of second class.

The solution of the equation $(i_X\omega_1=dh_1)_{|M_1}$ is given by
\[
X=q^3\frac{\partial}{\partial q^1}+q^4\frac{\partial}{\partial q^2}-q^4\frac{\partial}{\partial q^3}+(2q^3-q^2)\frac{\partial}{\partial q^4}-q^2\frac{\partial}{\partial p_3}
\]

We will study now the solutions of the Hamilton-Jacobi equation. So, we look for $\gamma\in \Lambda^1(\mathbb{R}^4)$ such that
\begin{enumerate}
\item $\gamma(\mathbb{R}^4)\subset M_1$
\item $d(h_1\circ \gamma)=0$
\item $d\gamma=0$
\end{enumerate}

If $\gamma(q^A)=(q^A,\gamma_1(q^A),\gamma_2(q^A),\gamma_3(q^A),\gamma_4(q^A))$, then the condition $\gamma(Q)\subset M_1$ gives
\[
\begin{array}{l}
\gamma_1(q^1,q^2,q^3,,q^4)=q^2+q^3 \\ \noalign{\medskip}
\gamma_2(q^1,q^2,q^3,,q^4)=0  \\ \noalign{\medskip}
\gamma_3(q^1,q^2,q^3,,q^4)=q^4 \\ \noalign{\medskip}
\gamma_4(q^1,q^2,q^3,,q^4)=0 \\ \noalign{\medskip}
\end{array}
\]
But $h_1\circ \gamma=-\frac{1}{2}((q^4)^2-2q^2q^3-(q^3)^2)$, so the equation $d(h_1\circ \gamma)=0$ if and only if $\gamma(q)=(q^A,0,0,0,0)$.

}
\end{example}

\begin{example}\label{exe}{\rm
This example has been discussed by K. Sundermeyer \cite{sundermeyer}. Let $L$ be the Lagrangian $L:T{\mathbb{R}}^2\rightarrow \mathbb{R} $  given by
\[
L(q^1,q^2,\dot{q}^1,\dot{q}^2)=\frac{1}{2}(\dot{q}^1)^2+ \dot{q}^2\, q^1+\dot{q}^2\, q^1.
\]

Then $FL$ is given by $FL:T{\mathbb{R}}^4\rightarrow T^*{\mathbb{R}}^4$
\[
FL(q^1,q^2,\dot{q}^1,\dot{q}^2)=(q^1,q^2,\dot{q}^1+q^2,q^1)
\]
and the primary constraints are
\[
\Phi^1(q^A,p_A)=p_2-q^1
\]
So
\[
\begin{array}{l}
M_1=\{(q^1,q^2,p_1,p_2)\in\mathbb{R}^4 \textrm{ such that } \ p_2=q^1\},
\end{array}
\]
and we can use $(q^1,q^2,p_1)$ as coordinates on $M_1$.

It follows that
\[\begin{array}{l}
E_L=\frac{1}{2}\dot{q}^1 \\  \noalign{\medskip}
h_1=\frac{1}{2}(p_1-q^2) \\ \noalign{\medskip}
\omega_1=dq^1\wedge dp^1+dq^2\wedge dq^1  \\ \noalign{\medskip}
\textrm{Ker}(\omega_1)=\displaystyle{\left<\frac{\partial }{\partial p_1}-\frac{\partial }{\partial q^2}\right>} \\
\end{array}
\]
Let
\[
h(q^A,p_A)=\frac{1}{2}(p_1-q^2)
\]
be an extension of the hamiltonian.

It is easy to see that at the points of $M_1:=\textrm{Im}(FL)$
\[
\{\Phi^1, h +u\Phi^1\}=0
\]
and therefore we have global dynamics.

The solution of the equation $(i_X\omega_1=dh_1)_{|M_1}$ is given by
\[
X=(p_1-q^2)\frac{\partial}{\partial q^1}+f\frac{\partial}{\partial q^2}+f\frac{\partial}{\partial p_1}+(p_1-q^2)\frac{\partial}{\partial p_2},
\]
where $f\in C^{\infty}(M_1)$

If we now look for $\gamma\in \Lambda^1(\mathbb{R}^4)$ such that
\begin{enumerate}
\item $\gamma(\mathbb{R}^4)\subset M_1$
\item $d(h_1\circ \gamma)=0$
\item $d\gamma=0$
\end{enumerate}
then $\gamma(q^1,q^2)=(q^1,q^2,\gamma_1(q^1,q^2),\gamma_2(q^1,q^2))$ given by
\[
\gamma(q^1,q^2)=(q^1,q^2,q^2,q^1)
\]
satisfies all the requiered conditions, because $p_1(\gamma(q^1,q^2)=q^2$, $\gamma=d(q^1\cdot q^2)$ and $h_1\circ \gamma(q^1,q^2)=\frac{1}{2}(q^2-q^2)=0$. Given an arbitrary  solution $X=(p_1-q^2)\frac{\partial}{\partial q^1}+f\frac{\partial}{\partial q^2}+f\frac{\partial}{\partial p_1}+(p_1-q^2)\frac{\partial}{\partial p_2}$ of the constarined dynamics, we have that
\[
X^{\gamma}=(f\circ \gamma)\frac{\partial}{\partial q^2}
\]
and also
\[
T\gamma(X^{\gamma})=(f\circ \gamma)\frac{\partial}{\partial q^2}+(f\circ \gamma)\frac{\partial}{\partial p^1}
\]
which is precisely  $X$ along $\textrm{Im}(\gamma)$.
}
\end{example}

\subsubsection{There are secondary constraints}\hfill

Next, we are going to describe several examples where secondary constraints appear.

\begin{example}{\rm
This example has been discussed by M.J. Gotay and J.M. Nester \cite{gotay3}. Let $L$ be the Lagrangian $L:T\mathbb{R}^2\rightarrow \mathbb{R} $  given by

\[
L(q^1,q^2,\dot{q}^1,\dot{q}^2)=\frac{1}{2}(\dot{q}^1)^2+q^2(q^1)^2.
\]

Then $FL$ is given by $FL:T\mathbb{R}^2\rightarrow T^*\mathbb{R}^2$
\[
FL(q^1,q^2,\dot{q}^1,\dot{q}^2)=(q^1,q^2,\dot{q}^1,0)
\]
and the primary constraints are
\[
\Phi^1(q^A,p_A)=p_2.
\]

So
\[
M_1=\{(q^1,q^2,p_1,p_2)\in\mathbb{R}^4 \textrm{ such that }  p_2=0\}
\]
and we can use $(q^1,q^2,p_1)$ as coordinates on $M_1$.

It follows that
\[\begin{array}{l}
E_L=\frac{1}{2}(\dot{q}^1)^2-q^2(q^1)^2 \\  \noalign{\medskip}
h_1=\frac{1}{2}(p_1)^2-q^2(q^1)^2\\  \noalign{\medskip}
\omega_1=dq^1\wedge dp_1  \\ \noalign{\medskip}
\textrm{Ker}(\omega_1)=\left<\frac{\partial}{\partial q^2} \right> \\
\end{array}
\]
Let
\[
h(q^A,p_A)=\frac{1}{2}(p_1)^2-q^2(q^1)^2
\]
be an arbitrary extension of the constarined hamiltonian $h_1$ to $T^*\R^2$.

It is easy to see that at the points of $M_1:=\textrm{Im}(FL)$ we have
\[\begin{array}{l}
\{\Phi^1, h +u^1\Phi^1\}=-(q^1)^2\\
\end{array}
\]
and therefore we need to restrict the dynamics adding a new constraint
\[
\Phi^2(q^A,p_A)=q^1.
\]
Now $M_2:=\{(q^1,q^2,p_1,p_2)\in\mathbb{R}^4 \textrm{ such that }  p_2=0, \ q^1=0\}$ and $Q_2:=\pi_Q(M_2)=\{(q^1,q^2)\in\mathbb{R}^2\ \textrm{ such that } q^1=0\} $. We have on $M_2$
\[\begin{array}{l}
\{\Phi^1, h +u^1\Phi^1\}=0\\
\{\Phi^2, h +u^1\Phi^1\}=p_1,
\end{array}
\]
and we need to restrict again the dynamics, adding the constraint
\[
\Phi^3(q^A,p_A)=p_1
\]

Now $M_3:=\{(q^1,q^2,p_1,p_2)\in\mathbb{R}^4 \textrm{ such that }  p_2=0, \ q^1=0, \ p^1=0\}$ and $Q_3=Q_2=\{(q^1,q^2)\in\mathbb{R}^2\ \textrm{ such that }  q^1=0\} $. Along $M_3$ we have
\[\begin{array}{l}
\{\Phi^1, h +u^1\Phi^1\}=0\\
\{\Phi^2, h +u^1\Phi^1\}=0\\
\{\Phi^3, h +u^1\Phi^1\}=0,
\end{array}
\]
and $M_3$ is the final constraint submanifold, $M_f$.
We can easily check that $\Phi^1$ is a first class constraint and $\Phi^2$, $\Phi^3$ are second class.

The solutions  of the equation $(i_X\omega_1=dh_1)_{|M_3}$ are of the form
\[
 X=f\frac{\partial}{\partial q^2},
\]
where $f\in C^{\infty}(M_3)$.

A solution of the Hamilton-Jacobi equation, should be $\gamma(q^1,q^2)=(q^1,q^2,\gamma_1(q^1,q^2),\gamma_2(q^1,q^2))$, such that
\begin{enumerate}
\item $\gamma(Q)\subset M_1$ and $\gamma_f(Q_f)\subset M_f$
\item $d(h_1\circ \gamma)_{|Q_f}=0$
\item $d\gamma=0$
\end{enumerate}

The condition $\gamma(Q)\subset M_1$ implies $\gamma_2=0$.
Next we compute $d\gamma$,
\[
d\gamma=\frac{\partial \gamma_1}{\partial q^2}dq^2\wedge dq^1+ \frac{\partial \gamma_2}{\partial q^1}dq^1\wedge dq^2=\frac{\partial \gamma_1}{\partial q^2}dq^2\wedge dq^1=0
\]
and we deduce that  $\frac{\partial \gamma_1}{\partial q^2}$ must vanish and $\gamma_1$ must be a function of $q^1$.

The condition $d(h_1\circ \gamma)_{|Q_f}=0$ can also  be easily computed. We have
\[
d(h_1\circ \gamma)=d(\frac{1}{2}(\gamma_1)^2-q^2(q^1)^2)=(\gamma_1\frac{\partial \gamma_1}{\partial q^1}-2q^2q^1)dq^1 + (q^1)^2dq^2
\]
and, along $Q_f=\{(q^1,q^2)\in\mathbb{R}^2\ \textrm{ such that }  q^1=0\}$, we deduce
\[
d(h_1\circ \gamma)_{|Q_f}=\gamma_1(0)\frac{\partial \gamma_1}{\partial q^1}(0) \, dq^1
\]

For example, if we take $\gamma_1=q^1$, $\gamma_2=0$, all the above conditions are satisfied, and $\gamma_f(Q_f)\subset M_f$.

Now, take a solution $X=f\frac{\partial}{\partial q^2}$; at the points of $Q_f$ we get
\[
X^{\gamma}(0,q^2)=(\pi_f)_*(f(0,q^2,0,0)\frac{\partial}{\partial q^2})= f(0,q^2,0,0)\frac{\partial}{\partial q^2}
\]
and so
\[
T\gamma_f(X^{\gamma}(0,q^2))=f(0,q^2,0,0)\frac{\partial}{\partial q^2},
\]
and we obtain the solution $X$ along $\textrm{Im}(\gamma_f)$.
}
\end{example}

\begin{example}{\rm
This example has been discussed by M.J. Gotay \cite{gotay4}. Let $Q:=\{(q^1,q^2)\in \mathbb{R}^2 \textrm{ such that } q^1\neq 0  \}$ and $L$ be the Lagrangian $L:TQ\rightarrow \mathbb{R} $  given by
\[
L(q^1,q^2,\dot{q}^1,\dot{q}^2)=\frac{1}{2q^1}(\dot{q}^2)^2.
\]

Then $FL$ is given by $FL:TQ\rightarrow T^*Q$
\[
FL(q^1,q^2,\dot{q}^1,\dot{q}^2)=(q^1,q^2,0,\dot{q}^2/q^1)
\]
and the primary constraints are
\[
\Phi^1(q^A,p_A)=p_1.
\]

So
\[
M_1=\{(q^1,q^2,p_1,p_2)\in TQ \textrm{ such that } p_1=0\}.
\]
and we can use $(q^1,q^2,p_2)$ as coordinates on $M_1$.

It follows that
\[\begin{array}{l}
E_L=L \\ \noalign{\medskip}
h_1(q^A,p_A)=\frac{q^1}{2}(p_2)^2 \\ \noalign{\medskip}
\omega_1=dq^2\wedge dp_2  \\ \noalign{\medskip}
\textrm{Ker}(\omega_1)=\left<\frac{\partial}{\partial q_1} \right> \\
\end{array}
\]
Let $h(q^A,p_A)=\frac{q^1}{2}(p_2)^2$
be an extension of the hamiltonian.

It is easy to see that at the points of $M_1:=\textrm{Im}(FL)$ we get
\[\begin{array}{l}
\{\Phi^1, h +u^1\Phi^1\}=-\frac{(p_2)^2}{2}\\
\end{array}
\]
and therefore we need to restrict the dynamics adding a new constraint
\[
\Phi^2(q^A,p_A)=p_2.
\]
Now $M_2:=\{(q^1,q^2,p_1,p_2)\in TQ \textrm{ such that } p_1=0, \ p_2=0\}$ and $Q_2:=\pi_Q(M_2)=Q $. At the points of $M_2$ we have
\[\begin{array}{l}
\{\Phi^1, h +u^1\Phi^1\}=0\\
\{\Phi^2, h +u^1\Phi^1\}=0,
\end{array}
\]
and $M_2$ is the final contraint manifold.
From $\{\Phi^1, \Phi^2\}=0$ we deduce that the constraints are all first class.

The solutions are of the form $X=f\frac{\partial}{\partial q^1}$ where $f\in C^{\infty}(M_2)$.

If we look for a solution of our Hamilton-Jacobi equation, $\gamma$, such that $\gamma(q^1,q^2)=(q^1,q^2,\gamma_1(q^1,q^2),\gamma_2(q^1,q^2))$, then the condition $\gamma(Q)\subset M_1$ implies $\gamma=0$ and $\gamma_f=0$. All  conditions are verified and, given a solution $X$, we obtain that $X^{\gamma}$ and $X$ are trivially $\gamma_f$-related.
}
\end{example}

\begin{example}{\rm
This example has been discussed by R. Skinner and R. Rusk \cite{ski}. Let $L$ be the Lagrangian $L:T\mathbb{R}^3\rightarrow \mathbb{R} $  given by
\[
L(q^1,q^2,q^3,\dot{q}^1,\dot{q}^2,\dot{q}^3)=\frac{1}{2}q^2(q^3)^2+\dot{q}^1\dot{q}^3.
\]

Then $FL:T\mathbb{R}^3\rightarrow T^*\mathbb{R}^3$ is given by
\[
FL(q^1,q^2,q^3,\dot{q}^1,\dot{q}^2,\dot{q}^3)=(q^1,q^2,q^3,\dot{q}^3,0,\dot{q}^1) \;,
\]
so that we have a primary constraint
$
\Phi^1(q^A,p_A)=p_2.
$
This means that the primary constraint submanifold is
\[
M_1=\{(q^1,q^2,q^3,p_1,p_2,p_3)\in\mathbb{R}^6 \textrm{ such that } p_2=0\},
\]
and then we can use $(q^1,q^2,q^3,p_1,p_3)$ as coordinates on $M_1$.

It follows that
\[\begin{array}{l}
E_L=\dot{q}^3\dot{q}^1+\dot{q}^1\dot{q}^3-L=-\frac{1}{2}q^2(q^3)^2+\dot{q}^1\dot{q}^3 \\ \noalign{\medskip}
h_1(q^A,p_A)=p_1p_3-\frac{1}{2}q^2(q^3)^2  \\ \noalign{\medskip}
\omega_1=dq^1\wedge dp_1+dq^3\wedge dp_3  \\ \noalign{\medskip}
\textrm{Ker}(\omega_1)=\left<\frac{\partial}{\partial q^2} \right>  .
\end{array}
\]
As in the previous cases, take an arbitrary extension of the hamiltonian $h_1$, for instance\[
h(q^A,p_A)=p_1p_3-\frac{1}{2}q^2(q^3)^2.
\]

It is easy to see that at the points of $M_1:=\textrm{Im}(FL)$
\[\begin{array}{l}
\{\Phi^1, h +u^1\Phi^1\}=\frac{1}{2}(q^3)^2\\
\end{array}
\]
and therefore we should restrict the dynamics adding a secondary constraint
\[
\Phi^2(q^A,p_A)=q^3.
\]
Now $M_2:=\{(q^1,q^2,q^3,p_1,p_2,p_3)\in\mathbb{R}^6 \textrm{ such that } p_2=0, \ q^3=0\}$ and $Q_2:=\pi_Q(M_2)=\{(q^1,q^2,q^3)\in\mathbb{R}^3 \textrm{ such that } \ q^3=0\} $. Along $M_2$, we have
\[
\begin{array}{l}
\{\Phi^1, h +u^1\Phi^1\}=0\\
\{\Phi^2, h +u^1\Phi^1\}=p_1.
\end{array}
\]
Therefore, we need again to restrict  the dynamics, adding the constraint
$
\Phi^3(q^A,p_A)=p_1\; .
$
Now $M_3:=\{(q^1,q^2,q^3,p_1,p_2,p_3)\in\mathbb{R}^6 \textrm{ such that } p_2=0, \ q^3=0, \ p_1=0\}$ and $Q_3=Q_2=\{(q^1,q^2,q^3)\in\mathbb{R}^3 \linebreak \textrm{such that} \ q^3=0\} $. Along $M_3$ we have
\[
\begin{array}{l}
\{\Phi^1, h +u^1\Phi^1\}=0\\
\{\Phi^2, h +u^1\Phi^1\}=0\\
\{\Phi^3, h +u^1\Phi^1\}=0
\end{array}
\]
and then $M_3$ is the final contraint manifold, denoted by $M_f$; therefore, $Q_f=Q_3$.
We deduce that the constraints are all first class.

The solutions of the equation $(i_X\omega_1=dh_1)_{|M_3}$ are of the form 
\[
X=p_3\frac{\partial}{\partial q^1}+f\frac{\partial}{\partial q^2},
\] 
where $f\in C^{\infty}(M_3)$.

Now we look for a solution of our Hamilton-Jacobi equation, that is  $\gamma(q^1,q^2,q^3)=(q^1,q^2,q^3,\gamma_1(q^1,q^2,q^3),\gamma_2(q^1,q^2,q^3),\gamma_3(q^1,q^2,q^3))$, such that
 \begin{enumerate}
 \item $\gamma(Q)\subset M_1$ and $\gamma_f(Q_f)\subset M_f$
 \item $d(h_1\circ \gamma)_{|Q_f}=0$
 \item $d\gamma=0$
 \end{enumerate}

The condition $\gamma(Q)\subset M_1$ implies $\gamma_2=0$; the condition $\gamma_f(Q_f)\subset M_f$ implies $(\gamma_f)_i
 =0$ for $i=1,2$ and, the condition $d(h_1\circ \gamma)=0$ is
 \[
 \begin{array}{l}
 d(h_1\circ \gamma)=\gamma_1\gamma_3-\frac{1}{2}q^2(q^3)^2\\
\noalign{\medskip}
=\left(\frac{\partial \gamma_1}{\partial q^1}\gamma_3 +\frac{\partial \gamma_3}{\partial q^1 }\gamma_1 \right)dq^1 +\left( \frac{\partial \gamma_1}{\partial q^2}\gamma_3  +\frac{\partial \gamma_3}{\partial q^2 }\gamma_1 + \frac{1}{2}(q^3)^2\right)dq^2 \\ \noalign{\medskip}+\left(\frac{\partial \gamma_1}{\partial q^3}\gamma_3 +\frac{\partial \gamma_3}{\partial q^3 }\gamma_1 +q^2q^3 \right)dq^3
\end{array}
\]
Hence,
\[
 \begin{array}{l}
d(h_1\circ \gamma)_{|Q_f}=\left(\frac{\partial \gamma_1}{\partial q^1}\gamma_3 +\frac{\partial \gamma_3}{\partial q^1 }\gamma_1\right)dq^1+ \left(\frac{\partial \gamma_1}{\partial q^2}\gamma_3 +\frac{\partial \gamma_3}{\partial q^2 }\gamma_1\right)dq^2 \\ \noalign{\medskip}+\left(\frac{\partial \gamma_1}{\partial q^3}\gamma_3 +\frac{\partial \gamma_3}{\partial q^3 }\gamma_1  \right)dq^3
\end{array}
\]
The condition $d\gamma=0$ implies
\[
\begin{array}{ll}
d\gamma= &\frac{\partial \gamma _1}{\partial q^2}dq^2\wedge dq^1 +\frac{\partial \gamma_1}{\partial q^3}dq^3\wedge dq^1 \\ \noalign{\medskip}
&+\frac{\partial \gamma_3}{\partial q^1}dq^1\wedge dq^3+\frac{\partial \gamma_3}{\partial q^2}dq^2\wedge dq^3=0
\end{array}
\]
taking into account that $\gamma_2=0$, 
and therefore
\[
\begin{array}{l}
\frac{\partial \gamma_1}{\partial q^2}=0 \\ \noalign{\medskip}
\frac{\partial \gamma_1}{\partial q^3}=\frac{\partial \gamma_3}{\partial q^1}\\ \noalign{\medskip}
\frac{\partial \gamma_3}{\partial q^2}=0
\end{array}
\]

 A particular solution is obtained putting $\gamma_1=\gamma_2=0$, and $\gamma_3$ an arbitrary function of $q^3$, for example $\gamma_3=q^3$.

For instance,  take $X=p_3\frac{\partial}{\partial q^1}+f\frac{\partial}{\partial q^2}$ and $S=\frac{1}{2}(q^3)^2$, then
\[
\gamma(q^1,q^2,q^3)=(q^1,q^2,q^3,0,0,q^3)
\]
and at the points of $Q_f$ we obtain
\[X^{\gamma}(q^1,q^2,0)=T\pi_f(0\frac{\partial}{\partial q^1}+f(q^1,q^2,0,0,0,0)\frac{\partial}{\partial q^2})=f(q^1,q^2,0,0,0,0)\frac{\partial}{\partial q^2},\]
so that
\[
T\gamma_f( X^{\gamma}(q^1,q^2,0))=f(q^1,q^2,0,0,0,0)\frac{\partial}{\partial q^2}
\]

We can also apply proposition \ref{112} to the latter example and obtain solutions of the extended equation.

For instance, consider $\gamma(q^1,q^2,q^3)=(q^1,q^2,q^3,\gamma_1(q^A),\gamma_2(q^A),\gamma_3(q^A))$ given by
\[
\gamma(q^1,q^2,q^3)=(q^1,q^2,q^3,q^3,0,q^1)
\]

We have
\begin{enumerate}
 \item  $\gamma_f(Q_f)\subset M_f$ because $$\gamma_f(Q_f)=\{(q^1,q^2,0,0,0,q^1)\in\mathbb{R}^6 \textrm{ such that } q^1, \, q^2 \in \mathbb{R}\}.$$
 \item  If we take coordinates $(q^1,q^2)$ in $Q_f$, then $d(h_1\circ \gamma_f)=d(0\cdot 0 -q^2 \cdot 0)=0$.
 \item $d\gamma=0$, in fact, $\gamma=d(q^1 \cdot q^3)$.
 \end{enumerate}

If we consider a solution $X=p_3\frac{\partial}{\partial q^1}+f\frac{\partial}{\partial q^2}$, we can compute
\[
X^{\gamma}(q^1,q^2,0)=T\tau_Q\circ X\circ \gamma_f(q^1,q^2,0)=q^1\frac{\partial}{\partial q^1}+f(q^1,q^2,0)\frac{\partial}{\partial q^2}
\]
and also
\[
T\gamma_f (X^{\gamma}(q^1,q^2,0))=q^1\frac{\partial}{\partial q^1}+f(q^1,q^2,0)\frac{\partial}{\partial q^2}+q^1\frac{\partial}{\partial p_ 3}
\]
which is a solution of the equation $i_X \, \omega_3=dh_3$ where, if $i_3:M_3\rightarrow T^*Q$ is the inclusion on $T^*Q$ and $j_3 :M_3\rightarrow M_1$, then $\omega_3=i_3^*(\omega_Q)$ and $h_3=j_ 3^*(h_1)$.

Note that $\gamma$ in this case is not a solution of our Hamilton-Jacobi problem because $d(h\circ \gamma)(q^1,q^2,0)=q^1dq^3\neq 0$
}
\end{example}

\subsection{Relation to classical Hamilton-Jacobi Theory}
In this section we will connect the Hamilton-Jacobi theory developed in the previous sections with the classical Hamilton Jacobi theory on $T^*Q$ using an appropriate extended hamiltonian.

 We will use the same notation that in section \ref{dirac}. We start with an almost regular lagrangian $L:TQ\rightarrow \mathbb{R}$, and then $\textrm{Im}(FL)=M_1$ is a differentiable submanifold of $T^* Q$ and, in addition, we can define $h_1$ implicitly by $h_1\circ FL=E_L$.
We denote $\omega_1=j_1^*\omega_Q$, where $j_1:M_1\rightarrow T^*Q$ is the inclusion and $\omega_Q$ is the canonical symplectic form of the cotangent bundle. We take local coordinates $(q^A,p^A)$ in an open set $U\subset T^*Q$, such that $M_1$ is given locally by the vanishing of independent functions $\Phi^i(q^A,p^A)$, called primary constraints.

Remember that the equations of motion have the form $(i_X\omega_1=dh_1)_{|U_1}$, where $U_1:=M_1\cap U$. This equations are equivalent to $(i_X\omega_Q=dh+\mu_i d\Phi^i)_{|U_1}$ where $h$ is any extension of $h_1$ to $U$ defined on $M_1$ and  $\mu_i$ are Lagrange multipliers.

\subsubsection{Case I: There are only primary constraints}
 First, we suppose that there exist a global solution $X$, i.e. $X$ is a vector field on $M_1$ that satisfies the equations of motion. We reorder constraint functions in two classes: first class constraints denoted by $\chi^a$ and second class constraints denoted by $\xi^b$. We also denote by $ u_a$ and $\lambda_b$ the corresponding Lagrange multipliers. Then the equations of motion are
\[
(i_X \, \omega_Q = dh+u_ad\chi^a+\lambda_b d\xi^b)_{|M_1}.
\]
Now, suppose that $u_a$ and $\lambda_b$ are functions defined on $U$. It is clear that $X$ is the restriction to $M_1$ of the hamiltonian vector field corresponding to a hamiltonian of the form $h+u_a\chi^a+\lambda_b\xi^b$. In fact, all the solutions of the equations of motion are obtained in this way varying the functions $u_a$ arbitrarily and with prescribed values of $\lambda_b$. Next, we are looking for a solution of our Hamilton-Jacobi problem, that is, a 1-form $\gamma$ satisfying
\begin{enumerate}
\item $d\gamma=0$
\item $\textrm{Im}(\gamma)\subset M_1$
\item $d(h_1\circ \gamma)=0$
\end{enumerate}
Condition (iii) can be easily checked that it is equivalent to $d((h+u_a\chi^a+\lambda_b\xi^b)\circ \gamma)=0$ because $(h+u^a\chi^a+\lambda^b\xi^b)_{|M_1}=h_1$. So, it is evident that the solutions of the classical Hamilton-Jacobi equation for the hamiltonians $h+u^a\chi^a+\lambda^b\xi^b$ (where $u^a$ are arbitrary functions and the rest are fixed)  inside $M_1$ and the solutions for our Hamilton-Jacobi problem coincide.

\subsubsection{Case II: The general case}
Suppose now that the algorithm do not stop at $M_1$, then we obtain the sequence of manifolds
 \[
\cdots M_k \hookrightarrow \cdots\hookrightarrow  M_2 \hookrightarrow M_ 1
\hookrightarrow T^*Q.
\]
and we suppose that the algorithm stabilizes in a manifold $M_f$ of dimension$>0$.

We can reorder the  constraints in first and second class (maybe changing the independent set of constrainsts). We will denote $\chi^a$ and $\xi^b$ the primary first and second class constraints and by $\psi^c$ and $\theta^d$  the secondary first and second class constraints. We will also denote by $u_a,\lambda_b,v_c$ and $w_d$ the corresponding Lagrange multipliers. Again a solution $X$ of the equations of motion verifies
\[
(i_X \, \omega_Q = dh+u_ad\chi^a+\lambda_bd\xi^b)_{|M_f}.
\]
As above, $X$ is the restriction to $M_f$ of the hamiltonian vector field given by the hamiltonian $h+u_a\chi^a+\lambda_b\xi^b$ where some multipliers are determined applying the constraint algorithm.

We are looking for $\gamma\in \Lambda^1(Q)$ satisfying
\begin{enumerate}
\item $d\gamma=0$
\item $\textrm{Im}(\gamma)\subset M_1$
\item $\gamma(Q_f)\subset M_f$
\item $d(h_1\circ \gamma_1)=0$
\end{enumerate}

Note that (iv) is  equivalent  to the equation $d((h+u_a\chi^a+\lambda_b\xi^b)\circ \gamma)=0$ because $(h+u_a\chi^a+\lambda_b\xi^b)_{|M_1}=h_1$, and so, the solutions of the classical Hamilton-Jacobi theory contained in $M_f$ for the hamiltonians $h+u_a\chi^a+\lambda_b\xi^b$ are just the solutions of our Hamilton-Jacobi problem.

\subsection{Relation to other theories}
The Hamilton-Jacobi theory for degenerate lagrangians have been discussed by several authors in the last 20 years. Let us recall some previous attempts.

\begin{enumerate}

\item In the papers by Longhi {\it et al.} \cite{gomis1,gomis2} it is discussed the case of a time independent lagrangian which is homogeneous in the velocities. It is shown that we can substitute an arbitrary lagrangian by an homogeneous one using the traditional procedure by adding new variables and, then, this new homogeneous lagrangian has zero energy. The authors show that the hamiltonian can be added as a new constraint and, in consequence,  they restrict themselves to the case when the hamiltonian is identically zero. The integrability condition for the resultant Hamilton-Jacobi equations implies that they can only consider first class constraints. On the other hand, in the paper by Rothe and F. G. Scholtz \cite{Rothe} an almost-regular  lagrangian $L(t,q^A,\dot{q}^A)$ is considered. If the Hessian ($\frac{\partial^2L}{\partial \dot{q}^A \partial \dot{q}^B}$), has rank $n-m_1$ then, the constraint submanifold $M_1$ is locally described by coordinates $(q^A, p_a)$, where  only , $a=m_1+1,\ldots,n$. The remaining momenta $p_\alpha$; $\alpha=1,\ldots,m_1$ are functions of $t$, $q^A$, $p_a$, that is,  $p_{\alpha}=-f_{\alpha}(t,q^A,p_a)$  and represent the primary constraints  $\phi_{\alpha}(t,q^A,p_A)=p_{\alpha}+f_{\alpha}(t, q^A,p_a)$. Then they consider the system of partial differential equations
\[
\begin{array}{l}
\frac{\partial S}{\partial t}+h_1(t,q^A,\frac{\partial S}{\partial q_a})=0 \\ \noalign{\medskip}
\frac{\partial S}{\partial q^{\alpha}}+ f_{\alpha}(t,q^A,\frac {\partial S}{\partial q_a})=0 \quad b=1,\ldots,m_1
\end{array}
\]
where $h_1$ is the hamiltonian defined on the primary constraint manifold by the projection of the lagrangian energy.

\item The theory discussed in  \cite{cari} is  similar to our theory in the case of  global dynamics, but they do not take into account secondary constraints. The authors also use the lagrangian homogeneous formalism to obtain the standard Hamilton-Jacobi theory for time dependent systems.

\item M. Leok and collaborators \cite{sosa} use the Dirac structures setting, and secondary constraints are not considered.

\end{enumerate}

\subsection{Lagrangian setting}

The equations of motion are globally expressed by the presymplectic equation
\begin{equation}\label{equ}
i_ {\xi}\,\omega_L=dE_L,
\end{equation}
where a possible solution $\xi$ is not in principle a SODE.

Therefore, in addition to the problem of finding solutions for \eqref{equ}, we must study the second order differential problem, that is, we shall obtain a solution of \eqref{equ} satisfying the additional condition $S\xi=\Delta$.

If we apply the constraint algorithm to the presymplectic system $(TQ,\omega_L,dE_L)$ we obtain a sequence of submanifolds.
\[
\cdots P_k \hookrightarrow \cdots \hookrightarrow P_2 \hookrightarrow P_ 1:= TQ
\]
Assume that the algorithm stabilizes at some $P_{k+1}=P_k=P_f$, which is the final constraint submanifold.

If we consider, as above, the presymplectic system $(M_1,\omega_1,dh_1)$, and apply the constraint algorithm to the equation
\begin{equation}\label{equ2}
i_{X}\, \omega_1=dh_1
\end{equation}
we obtain a sequence of submanifolds
\[
\cdots M_k \hookrightarrow \cdots \hookrightarrow M_2 \hookrightarrow M_ 1
\hookrightarrow T^*Q.
\]
such that
\[ FL(P_i)=M_i \textrm{, for any $i$, }
\]
and
\[
FL_i:=FL_{|P_i}:P_i\rightarrow M_i
\]
are surjective submersions.

As a consequence, both algorithms stabilizes at the same step, say $k$, and then
\[
FL(P_f)=M_f
\]
and
\[
FL_f:P_f\rightarrow M_f
\]
is a surjective submersion.
Moreover, we have the following results.
\begin{proposition}
If $\xi$ is a $FL_f$-projectable solution of \eqref{equ}, then its projection $TFL_f(\xi)$ is a solution of \eqref{equ2}.

Conversely, if $X$ is a solution of \eqref{equ2}, then any $FL_f$ projectable vector field on $P_f$ which projects on $X$, is a solution of \eqref{equ}.
\end{proposition}

Next, we shall discuss the SODE problem as it was stated by M. J. Gotay and J. Nester \cite{gotaythesis,gotay2} (see
\cite{cari0} for an alternative description).

The results in \cite{gotaythesis,gotay2} can be summarized in the following result.

\begin{theorem}\label{SODE}\hfill
\begin{enumerate}

\item If $\xi$ is a $FL_f$-projectable vector field on $P_f$ then for any  $p\in M_f$  there exists a unique point in each fiber $FL_f^{-1}(p)$, denoted by $\eta_{\xi}(p)$ at which $\xi$ is a SODE. The point $\eta_{\xi}(p)$ is given by
\[
\eta_{\xi}(p):=T\tau_Q(\xi(p))
\]
\item The map
\[
\begin{array}{rccl}
\beta_{\xi}:& M_f &\longrightarrow & P_f \\ \noalign{\medskip}
& p&\rightarrow & \beta_{\xi}(p):= \eta_{\xi}(p)
\end{array}
\]
is a section of $FL_f:P_f\rightarrow M_f$ and on $\textrm{Im}(\beta_{\xi})$ there exists a unique vector field, denoted by $X_{\xi}$, which simultaneously satisfies the equations
\[
\begin{array}{l}
i_{X_{\xi}} \, \omega_L=dE_L\\ \noalign{\medskip}
SX_{\xi}=\Delta
\end{array}
\]
\end{enumerate}
\end{theorem}

We will now recall the construction of  a solution of the dynamical equation which simultaneously satisfies the SODE condition. If $X:=(FL_f)_{*}(\xi)$, then $X$ is a vector field on $M_f$ satisfying $i_{X} \, \omega_1=dh_1$. The vector field $X_{\xi}$ described in \textrm{(ii)} is given by
\[
X_{\xi}(\beta_{\xi}(p))=T\beta_{\xi}(X(p))
\]

A detailed proof can be seen in \cite{gotaythesis,gotay2}, but for the sake of completness, we recall here the way to choose the points on the fibers as it is stated in the Theorem \ref{SODE} (i).

In the last part of this section we come back to the Hamilton-Jacobi problem, but now in the lagrangian setting.

The application of the constraint algorithm is summarized in the following diagram
\[
\xymatrix{
P_1=TQ\ar[rr]^{FL} \ar[rrd]^{FL_1}  && T^*Q\\
P_2 \ar[u]^{g_2} \ar[rrd]^{FL_2}&& M_1 \ar[u]^{j_1}\\
\vdots & & M_2\ar[u]^{j_2} \\
P_f\ar[u]^{g_f}\ar[rrd]^{FL_f} && \vdots\\
&& M_f\ar[u]^{j_f}
}
\]

Assume, as before, that $Q_i=\pi_Q(M_i)$ are submanifolds and $\pi_i={\pi_Q}_{|M_i}:M_i\rightarrow Q_i$ are surjective submersions. Since $\tau_Q=\pi_Q\circ FL$, then $\tau_Q(P_i)=\pi_Q(M_i)=Q_i$, and $P_i$ also projects onto $Q_i$. We denote $\tau_f={\tau_Q}_{|P_f}:P_f\rightarrow Q_f$.

In consequence, the following diagram is commutative.
\[
\xymatrix{
P_f \ar[ddr]^{\tau_f}\ar[rrd]^{FL_f} && \\
                  && M_f \ar[dl]^{\pi_f}\\
&Q_f&
}
\]

Now, if $X$ is a solution of $i_{X}\,\omega_1=dh_1$ on $M_f$ and $\gamma$ is a $1$-form which is a solution of the Hamilton-Jacobi problem, that is,

\begin{enumerate}
\item $\gamma(Q)\subset M_1$ and $\gamma_f(Q_f)\subset M_f$
\item $d(h_1\circ \gamma_1)_{|Q_f}=0$
\item $d\gamma=0$
\end{enumerate}
then we can define $X^{\gamma}=T\pi_f\circ X\circ \gamma_f$. From Proposition \ref{112} we deduce that $X$ and $X^{\gamma}$ are $\gamma_f$-related.

On the other hand we can construct a $FL_f$-projectable vector field $\xi$ on $P_f$ which projects on $X$. Next we can apply Proposition \ref{SODE} and obtain the section $\beta_{\xi}:M_f\rightarrow P_f$. Recall that $X_{\xi}(\beta_{\xi}(p))=T\beta_{\xi}(X(p))$ is the unique vector field on $\textrm{Im}(\beta_{\xi})$ which satisfies the SODE condition and the equation $i_{X_{\xi}} \, \omega_L=dE_L$. The following lemma gives the relation between $\textrm{Im}(\beta_{\xi})$ and $Q_f$.
\begin{lemma} $\textrm{Im}(\beta_{\xi})$ is a submanifold of $ TQ_f$.\end{lemma}

{\bf Proof:} Since $X_{\xi}$ verifies the SODE condition, then
\[
T\tau_Q(X_{\xi}(p))=\tau_{TQ}(X_{\xi}(p))
\]
for any $p\in \textrm{Im}(\beta_{\xi})$.

Since $X_{\xi}$ is tangent to $\textrm{Im}(\beta_{\xi})$, and since $\textrm{Im}(\beta_{\xi})$ is a submanifold of $ P_f$ and $\tau_Q(P_f)=Q_f$, then $T\tau_Q(X_{\xi}(p))\in TQ_f$.

On the other hand $\tau_{TQ}(X_{\xi}(p))=p\in \textrm{Im}(\beta_{\xi})$, and using the SODE condition we deduce that $p\in TQ_f$.\hfill $\Box$

Remember that $X_{\xi}$ and $X$ are $\beta_{\xi}$-related and $X$ and $X^{\gamma}$ are $\gamma_f$-related, so we deduce that $X_{\xi}$ and $X^{\gamma}$ are $\beta_{\xi}\circ \gamma_f$-related too. Moreover, since $X_{\xi}$ satisfies the SODE condition, we can find a better description of $\beta_{\xi}\circ \gamma_f$.
\begin{proposition} We have  \[ \beta_{\xi}\circ \gamma_f=X^{\gamma}.\]

\end{proposition}
{\bf Proof}
Since $X_{\xi}$ verifies the SODE condition, then given $q\in Q_f$ we obtain
\[
T\tau_q(X_{\xi}(\beta_{\xi}\circ\gamma_f(q)))=\tau_{TQ}(X_{\xi}(\beta_{\xi}\circ\gamma_f(q))).
\]
Therefore,
\[
T\tau_Q(X_{\xi}(\beta_{\xi}\circ\gamma_f(q)))=T\tau_Q\circ T\beta_{\xi}(X(q))=T\pi(X(\gamma(q)))=X^{\gamma}(q)
\]
where we have used that $\tau_Q=\pi\circ FL$ and $FL\circ \beta_{\xi}=id_{M_f}$.

On the other hand,
\[
\tau_{TQ}(X_{\xi}(\beta_{\xi}\circ\gamma_f(q)))=(\beta_{\xi}\circ\gamma_f(q))
\]
Then, using the SODE condition we get $X^{\gamma}=\beta_{\xi}\circ\gamma_f$.
\hfill $\Box$

The following corollary is immediate.

\begin{corollary} The vector fields $X_{\xi}$ and $X^{\gamma}$ are $X^{\gamma}$-related, i.e.
\[
X_{\xi}(\beta_{\xi}\circ \gamma_f(q))=TX^{\gamma}(X^{\gamma}(q))
\]
or equivalently
\[
X_{\xi}(\beta_{\xi}\circ \gamma_f(q))=(X^{\gamma})^{C}(X^{\gamma}(q)),
\]
where $(X^{\gamma})^C$ denotes the complete lift of the vector field $X^{\gamma}$.
\end{corollary}

\subsection{Example: Lagrangian setting}
\begin{example}{\rm
We will revisite example \ref{exe} and discuss the Hamilton-Jacobi problem for the Euler-Lagrange equation. The lagrangian function is
\[
L(q^1,q^2,\dot{q}^1,\dot{q}^2)=\frac{1}{2}(\dot{q}^1)^2+ \dot{q^2}\, q^1+\dot{q^1}\, q^1.
\]

Then $FL$ was given
\[
FL(q^1,q^2,\dot{q}^1,\dot{q}^2)=(q^1,q^2,\dot{q}^1+q^2,q^1)
\]
and the primary constraints are
\[
\Phi^1(q^A,p_A)=p_2-q^1
\]
So
\[
\begin{array}{l}
M_1=\{(q^1,q^2,p_1,p_2)\in\mathbb{R}^4 \textrm{ such that } \ p_2=q^1\}.
\end{array}
\]
and we can use $(q^1,q^2,p_1)$ as coordinates on $M_1$.

It follows that
\[\begin{array}{l}
E_L=\frac{1}{2}\dot{q}^1 \\  \noalign{\medskip}
h_1=\frac{1}{2}(p_1-q^2) \\ \noalign{\medskip}
\omega_1=dq^1\wedge dp^1+dq^2\wedge dq^1  \\ \noalign{\medskip}
\textrm{Ker}(\omega_1)=\left<\frac{\partial }{\partial p_1}-\frac{\partial }{\partial q^2}\right> \\
\end{array}
\]
Let
\[
h(q^A,p_A)=\frac{1}{2}(p_1-q^2)
\]
be an extension of the hamiltonian $h_1$.

It is easy to see that, at the points of $M_1:=\textrm{Im}(FL)$, we have
\[
\{\Phi^1, h +u\Phi^1\}=0
\]
and therefore we are in presence of global dynamics

The solution of the equation $(i_X\omega_1=dh_1)_{|M_1}$ is given by
\[
X=(p_1-q^2)\frac{\partial}{\partial q^1}+f\frac{\partial}{\partial q^2}+f\frac{\partial}{\partial p1}+(p_1-q^2)\frac{\partial}{\partial p^2},
\]
where $f\in C^{\infty}(M_1)$

Recall also, that a solution of our Hamilton-Jacobi problem,  $\gamma(q^1,q^2)$ $=(q^1,q^2,\gamma_1(q^1,q^2),\gamma_2(q^1,q^2))$, was given by
\[
\gamma(q^1,q^2)=(q^1,q^2,q^2,q^1)
\]
If we take a solution $X=(p_1-q^2)\frac{\partial}{\partial q^1}+f\frac{\partial}{\partial q^2}+f\frac{\partial}{\partial p_1}+(p_1-q^2)\frac{\partial}{\partial p_2}$ we  can compute
\[
X^{\gamma}=0\frac{\partial}{\partial q^1}+f\frac{\partial}{\partial q^2}
\]
and also
\[
(X^{\gamma})^C(q^1,q^2,\dot{q^1},\dot{q^2})=(f\circ\gamma)\frac{\partial}{\partial q_1}+\left((\frac{\partial (f\circ \gamma)}{\partial q^1})\dot{q}^1+(\frac{\partial (f\circ \gamma)}{\partial q^2})\dot{q}^2\right)\frac{\partial}{\partial \dot{q}^2},
\]
then
\[
(X^{\gamma})^C(X^{\gamma})=(X^{\gamma})^C(q^1,q^2,0,(f\circ \gamma)(q^1,q^2))=(f\circ\gamma)\frac{\partial}{\partial q_1}+(\frac{\partial (f\circ \gamma)}{\partial q^2})\dot{q}^2\frac{\partial}{\partial \dot{q}^2}.
\]

This vector field along $X^{\gamma}$ satisfies the SODE condition. We can consider now the equation $i_ {\xi}\, \omega_L=dE_L$
\[
\begin{array}{ll}
\omega_L&=d(\frac{\partial L}{\partial \dot{q}^1}dq^1+\frac{\partial L}{\partial \dot{q}^2}dq^2)=d((\dot{q}^1+q^2)dq^1+q^1dq^2)\\ \noalign{\medskip} &=d\dot{q}^1\wedge dq^1+dq^2\wedge dq^1+dq^1\wedge dq^2=d\dot{q}^1\wedge dq^1.
\end{array}
\]
So, $i_{(X^{\gamma})^C(X^{\gamma})}\, \omega_L=0$ and $dE_L(X^{\gamma})=\dot{q}^1d\dot{q^1}(X^{\gamma})=0$ and thus $$i_{(X^{\gamma})^C(X^{\gamma})}\, \omega_L=dE_L(X^{\gamma}).$$Therefore $(X^{\gamma})^C(X^{\gamma})$ satisfies Euler-Lagrange equations and the SODE condition.

}
\end{example}

\end{document}